\documentclass[acmsmall]{acmart}
\AtBeginDocument{%
	\providecommand\BibTeX{{%
			\normalfont B\kern-0.5em{\scshape i\kern-0.25em b}\kern-0.8em\TeX}}}

\setcopyright{none}
\acmDOI{.}

\usepackage{paralist}
\usepackage{xcolor}
\usepackage[caption=false]{subfig}
\usepackage{soul}

\def\itemrange#1{%
  \addtocounter{enumi}{1}%
  \edef\labelenumi{\theenumi--\noexpand\theenumi}%
  \addtocounter{enumi}{-1}%
  \addtocounter{enumi}{#1}%
  \item
\def\labelenumi{\theenumi}}

\begin{document}
\title{Predicting traffic overflows on private peering}

\author{Elad Rapaport}
\email{eladr@post.bgu.ac.il}
\affiliation{%
	\institution{Telekom Innovation Laboratories and Department of Software and Information Systems Engineering}
	\institution{ Ben-Gurion University of the Negev}
	\city{Beer-Sheva}
	\country{Israel}
}

\author{Ingmar Poese}
\email{ingmar@benocs.com}
\affiliation{%
	\institution{BENOCS GmbH}
	\city{Berlin}
	\country{Germany}
}

\author{Polina Zilberman}
\email{polinaz@bgu.ac.il}
\affiliation{%
	\institution{Telekom Innovation Laboratories}
	\institution{Ben-Gurion University of the Negev}
	\city{Beer-Sheva}
	\country{Israel}
}

\author{Oliver Holschke}
\email{oliver.holschke@telekom.de}
\affiliation{%
	\institution{Telekom Innovation Laboratories}
	\city{Berlin}
	\country{Germany}
}

\author{Rami Puzis}
\email{puzis@bgu.ac.il}
\affiliation{%
	\institution{Telekom Innovation Laboratories and Department of Software and Information Systems Engineering}
	\institution{ Ben-Gurion University of the Negev}
	\city{Beer-Sheva}
	\country{Israel}
}

\renewcommand{\shortauthors}{Puzis, et al.}

\newcommand{\todo}[1]{\textit{\textcolor{red}{TODO: #1}}}

\newcommand{\avg}{\texttt{AVG}}
\newcommand{\std}{\texttt{STD}}

\begin{abstract}
\normalsize
Large content providers and content distribution network operators usually connect with large Internet service providers (eyeball networks) through dedicated private peering. The capacity of these private network interconnects is provisioned to match the volume of the real content demand by the users. Unfortunately, in case of a surge in traffic demand, for example due to a content trending in a certain country, the capacity of the private interconnect may deplete and the content provider/distributor would have to reroute the excess traffic through transit providers. Although, such overflow events are rare, they have significant negative impacts on content providers, Internet service providers, and end-users. These include unexpected delays and disruptions reducing the user experience quality, as well as direct costs paid by the Internet service provider to the transit providers. 
If the traffic overflow events could be predicted, the Internet service providers would be able to influence the routes chosen for the excess traffic to reduce the costs and increase user experience quality. In this article we propose a method based on an ensemble of deep learning models to predict overflow events over a short term horizon of 2-6 hours and predict the specific interconnections that will ingress the overflow traffic. The method was evaluated with 2.5 years' traffic measurement data from a large European Internet service provider resulting in a true-positive rate of 0.8 while maintaining a 0.05 false-positive rate. 
The lockdown imposed by the COVID-19 pandemic reduced the overflow prediction accuracy. Nevertheless, starting from the end of April 2020 with the gradual lockdown release, the old models trained before the pandemic perform equally well. 

\end{abstract}

\maketitle
\vspace{-0.2cm}
\section{Introduction}
\label{intro}

\begin{figure}[b!]
\centering
\includegraphics[width=8cm]{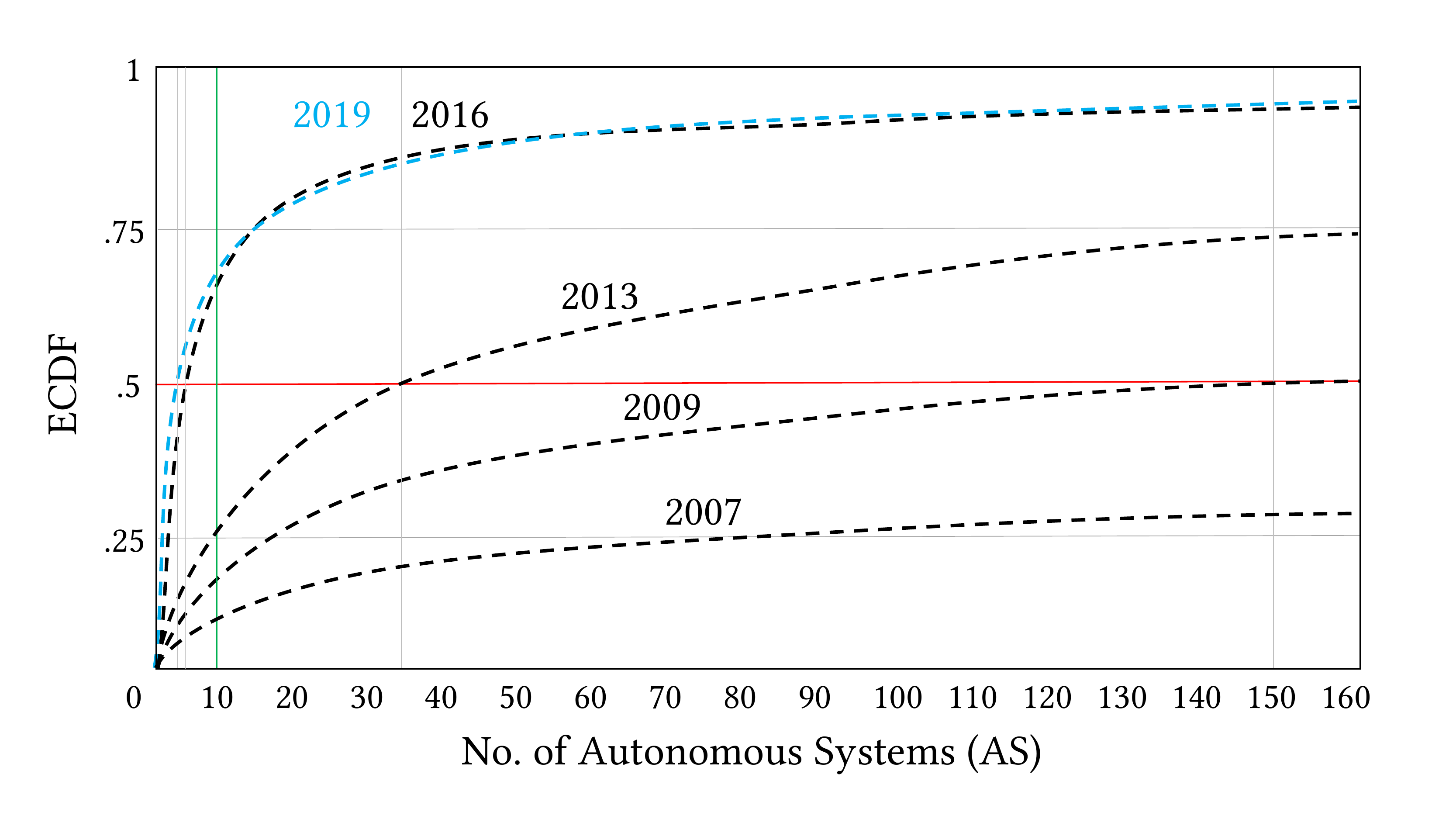}
\caption{Development of cumulative shares of Internet traffic distributed by ASes over past 13 years. The curves for 2007 to 2016 are from \cite{Labovitz2010,Labovitz2016}, for 2019 the underlying data is from the data set used in this study.}
\label{fig:CDF2}
\end{figure}

\begin{figure}
\centering
\includegraphics[width=8cm]{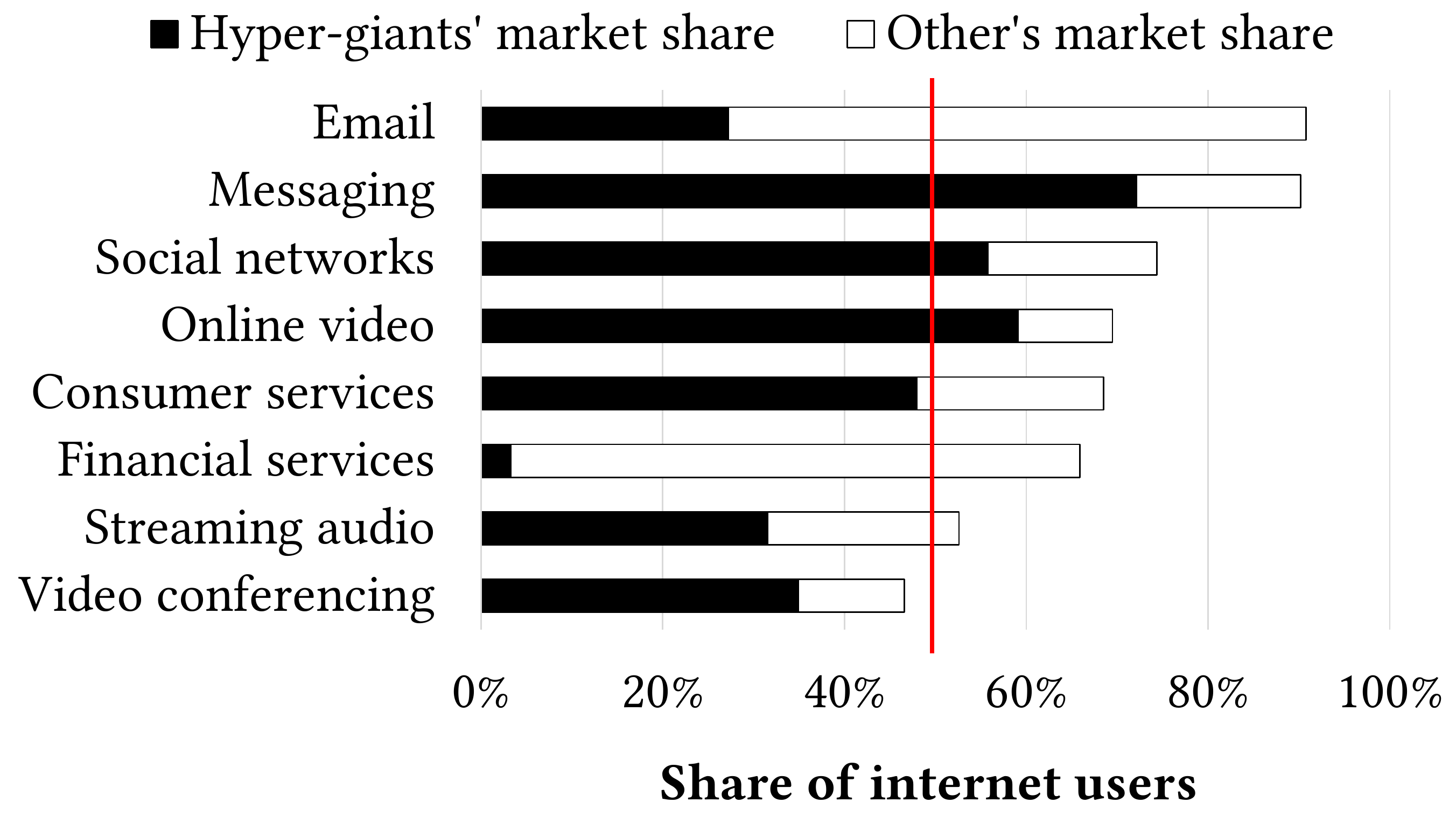}
\caption{Indicative usage of Internet applications and market shares of hyper-giants. The estimation is based on statistics on online activities of adults in the US from 2017 \cite{statista2017} and our data on hyper-giant prevalence.}
\label{fig:stacked-bars}
\end{figure}

Most of the Internet traffic consumed by end-users today is produced by a few largest content providers and distributors.  
This is a result of a process that has taken place over the past decade leading to a situation where ten Autonomous Systems (AS) produce more than 60\% of total traffic to networks, as depicted in Fig.~\ref{fig:CDF2} \cite{Labovitz2010,Labovitz2016}.
We refer to the traffic produced by the major ASes as the \textit{hyper-giant traffic}. 
The increased volume of hyper-giant traffic correlates is a result of an increased number of end-users consuming the services that are directly related to this traffic, see Fig.~\ref{fig:stacked-bars}.
The share of Internet users in the United States in 2013 that were accessing online videos accounted for 75.7\%.
This share is constantly increasing and is projected to reach 83.3\% in 2020.
The subscriber base of Netflix in the United States has doubled from approximately 30 million subscribers in 2013 to approximately 60 million subscribers in 2019.
Netflix is considered to be among the top ten hyper-giant ASes in many countries.
Similar significant growth rates in customer bases of hyper-giants can be observed in Central Europe as well.


The concentration of traffic on a few hyper-giants has significant effects on the Internet infrastructure. 
In order to optimize their performance, most hyper-giants setup and operate private network interfaces (PNI) with the networks providing Internet services to the end-users (also known as, eyeball networks).
PNIs are dedicated to carrying only the traffic emanating from the hyper-giant toward the end-users~\cite{wohlfart2018leveraging}. 
Due to the use of PNIs by hyper-giants, 
any disruption of traffic typically delivered through a PNI 
negatively impacts a larger fraction of the end-users and a larger portfolio of services.  
The degraded customer experience due to traffic shift from PNIs now results in 
a greater chance of customer churn, 
more penalties due to SLA (service level agreement) violations, 
and a negative impact on brand perception. 
The total value-at-risk by hyper-giant traffic disruptions 
has risen significantly \cite{Cholda2013}.

PNIs deliver a substantial volume of the consumed traffic directly into the eyeball network, which reduces the need to upgrade general interconnection points with public peering networks. 
Unfortunately, a hyper-giant may experience a mapping issue, flash crowd, PNI link failure, or any situation when the capacity of the PNI between the hyper-giant and the eyeball network is exhausted. 
To prevent the loss of traffic,
it falls back to deliver the traffic to the eyeball network through the neglected public peering networks.
We refer to such fall-backs 
as \textit{overflow events}.

Since public peering interconnections are not well built for asymmetric traffic volumes, 
overflow events can overload the public peering interconnections and result in service degradation not only for the specific hyper-giant but also for all content providers that do not maintain their own PNIs. 
The main consequence of overflow events is, thus, lost traffic and degraded service quality.

Timely and accurate prediction of overflow events would enable the eyeball network providers to respond to such events with countermeasures such as 
requesting the hyper-giant to reroute the traffic 
or remap the users to different clusters in a way that will cause fewer infrastructure disruptions.     
Predicting network traffic and link failures is extremely challenging in a general case~\cite{kodialam2006throughput}.
However, user Internet consumption is largely predictable. 
For example, deep learning was shown be effective in cellular and wireless traffic prediction tasks~\cite{huang2017study,zhang2018cellular,zhang2019DL-survey}.  
As we show in this paper, traffic surge from hyper-giants is largely predictable as well. 


Overflow events happen only during peak hours between 17:00-21:00. 
In order to allow a reasonable time for reaction, the prediction should be made at least two hours before the actual event.  
In this article, we (1) define the overflow prediction problem and (2) propose an ensemble of deep learning models to predict daily overflow events. The models:
\begin{compactitem}
    \item[(3)] predict overflow events with a true-positive rate of 0.8, and false-positive rate of 0.05; 
    \item[(4)] pin-point the specific handovers which will carry the overflow traffic;
    \item[(5)] survived the COVID-19 pandemic restoring the accuracy after lockdown release despite being six months old.   
\end{compactitem}
Finally, (6) analysis of the trained models using explainable artificial intelligence (XAI) techniques, shows that the most significant clues for overflow events are found in the traffic on the same day a few hours before the event and on the same weekday a week before the event.





\vspace{-0.2cm}
\section{Related work}
\label{sec:related}
\subsection{Deep learning (DL) for network traffic prediction}
Network traffic prediction, a form of multivariate time-series (MTS) prediction, is a challenging task that involves the leveraging of inter-dependencies between different variables over time, in order to make predictions for a specific variable or a set of variables in the future.

Vector auto-regression (VAR) is one of the most simple and common methods for MTS prediction~\cite{zivot2006vector}. It is a linear model, where each predicted variable is a weighted sum of past measurements (of itself and other variables in the time-series). Linear models usually fail to capture complex intricacies of large MTS datasets, and this has been proved by the following related works.

Yu and Chen~\cite{yu1993traffic} implemented a simple 3-layer feed-forward neural network to predict video traffic in an ISDN network. They have shown that their model is superior to linear based models, such as autoregressive integrated moving average (ARIMA).


Mozo et al.~\cite{mozo2018forecasting} proposed a convolutional neural network architecture for predicting short-term changes in the amount of traffic crossing a data center network. 
Their model receives a single data stream at a time, sampled at different frequencies. Each sampling is fed into a different 1D convolutional filter and the outputs of these filters are summed. 


Andreoletti et al.~\cite{andreoletti2019network} applied Diffusion convolutional recurrent neural networks for network traffic prediction using multivariate data. They treated a specific backbone network as a graph, with each node being a unit in the network and each edge representing the bits transferred between two units in a certain time period. Their CNN is able to take advantage of the graph structure when predicting the traffic amounts for the next time period, outperforming methods such as long-short term memory (LSTM) units and multi-layer perceptrons.

Le et al. \cite{le2019deep} approached the problem of traffic matrix prediction by using convolutional LSTM networks, which apply recurrent operations to data that has been processed by convolution. They used a bi-directional recurrent neural network to impute missing data, by learning from future data in the time window fed to the predictor. Data imputation is helpful if the process of network data collection is resource consuming and error-prone, making it is infeasible to collect data indefinitely. In our approach all the data is ground-truth and the need for imputation is diminished.

Similar to the works above, we show that by using a CNN-based ensemble we are able to capture complex inter-dependencies of traffic from several sources and destinations.

\vspace{-0.2cm}
\subsection{Network traffic routing}
Routing as well as \texttt{traffic engineering} (TE) \cite{wang2006COPE,wang2008internetTE,benson2011} is may be internal or external. 
When it comes to internal traffic engineering, ISPs have free reign on how to implement this. However, external TE is usually restricted to border gateway protocol (BGP) announcements, which are either coarse-grained or produce much address space disaggregation \cite{feamster2003guidelines}.
In any case, TE aims to distribute the load into and inside the network such that overloads are avoided.

Overflow traffic is defined as traffic that shifts from\\ \texttt{Hyper-giant Private Network Interfaces} (PNI) onto general Internet peering links. 
Since most ingress traffic today enters eyeball networks via PNIs \cite{pujol2019}, the public peerings transfer less traffic because they do not carry the bulk load of the content anymore
. 
Thus, a fraction of rerouted PNI traffic can overload a public peering link. 
This in turn affects the traffic from the hyper-giant as well as the performance for all other sources utilizing the interconnection. 

For a network provider to react to an overflow event before performance degrades two steps are required. First, before the overflow happens, there needs to be a notification that an overflow is imminent. This notification can, in classical scenarios, be done via email and/or phone call between the hyper-giant and the ISP, but more often than not a more automated solution is required. We define step one to be an automated detection of possible overflow events to happen in the near future based \textbf{solely} on data available from the ISP itself. 

When step one detects an imminent overflow, a reaction must follow. We find step two to be the challenge of reacting \textbf{in time} to an overflow event to mitigate its effect on network performance. 

\vspace{-0.2cm}
\section{Terminology}
\label{sec:terminology}

\begin{figure}[t]
\centering
\includegraphics[width=\textwidth]{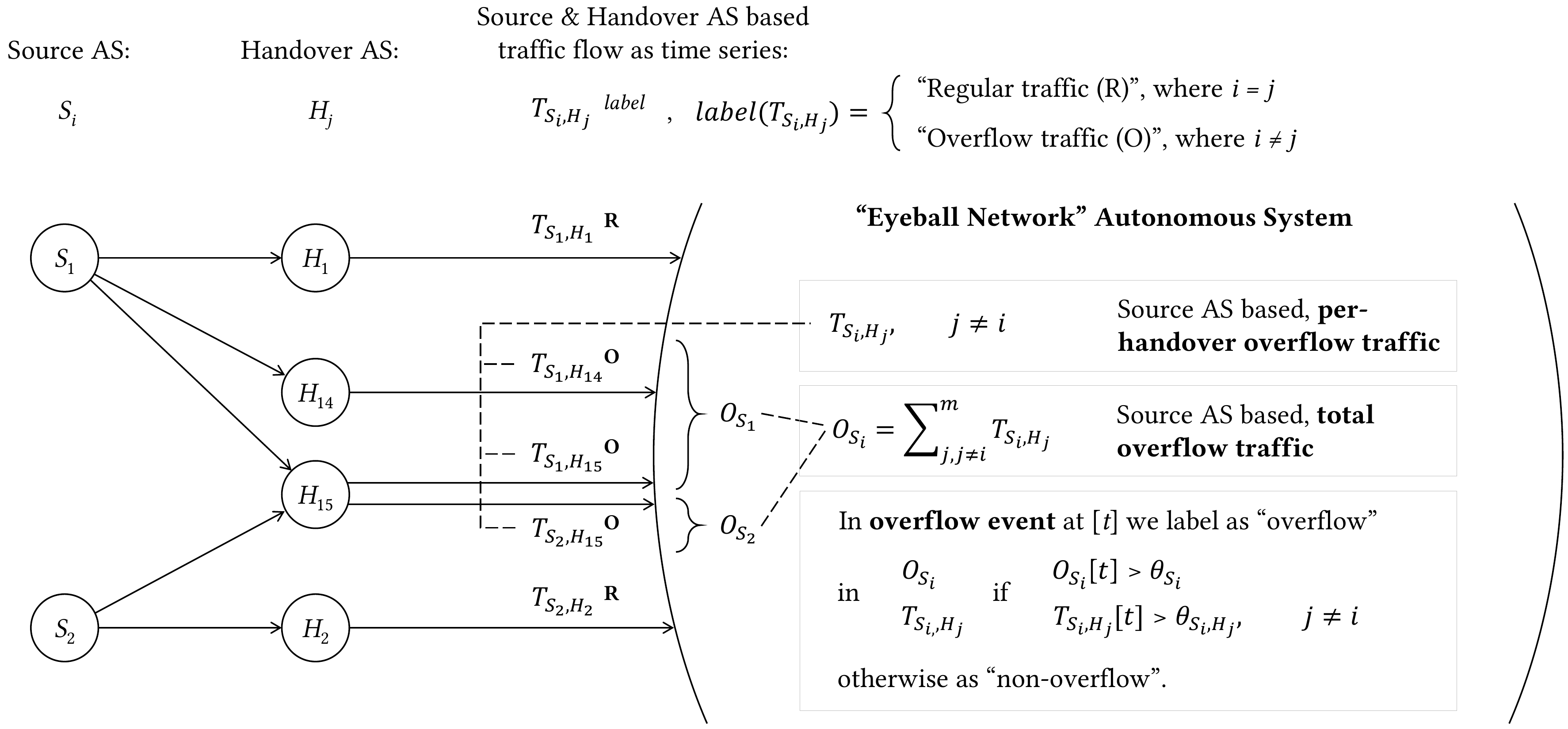}
\caption{Schematic overview of Source AS based, total \textit{overflow traffic}  -- in reference to the ``Eyeball Network'' AS -- as time-series $O_{S_{1}}$ and $O_{S_{2}}$ (here exemplary for two Source ASes $S_{1}$ and $S_{2}$, employing two exemplary Handover ASes $H_{14}$ and $H_{15}$) and their partial overflow traffic time-series $T_{S_{i},H_{j}}, j \neq i$.
They are further labeled as \textit{overflow} at $t$ when they exceed threshold $\theta_{S_{i}}$ and $\theta_{S_{i},H_{j}}$, respectively -- otherwise as \textit{non-overflow}.}
\label{fig:schematic-overview-2}
\vspace{-0.5cm}
\end{figure}

In this section, we outline the essential concepts and relationships representing \textbf{overflow traffic} and \textbf{overflow event} that will be used throughout the paper.
We will mark important terms used outside this section with boldface. 
Figure \ref{fig:schematic-overview-2} displays a schematic overview of the concepts.

\textbf{Autonomous System (AS)} is part of the Internet infrastructure, owned by an ISP \cite{rfc1930}.
\textbf{Eyeball network} is commonly used to refer to an AS through which end users look at the Internet. 
In this study, the vantage point in which overflow traffic and events are observed is an eyeball network.
The following concepts are defined form the perspective of an eyeball network in order to clearly convey topological positions and traffic directions.

\textbf{Source AS} from the perspective of the eyeball network is any AS which is the origin of the traffic consumed by eyeball network's end users. 
We denote source ASes as $S_1,S_2,\ldots,S_i,\ldots$, where $i$ identifies the AS. 
We study overflow traffic originating from source ASes that can have a significant impact on the eyeball network. 
Following the definition of \textbf{hyper-giant} by Pujol et al.~\cite{pujol2019} we focus in this research on source ASes that (1) produce at least 1\% of the total traffic delivered to broadband customers of the eyeball network, (2) publicly identify themselves as a content delivery network (CDN) or are registered as a ``content'' or ``enterprise network'' in the database PeeringDB\footnote{\url{https://www.peeringdb.com/}} and (3) the majority of their traffic (>60\%) is transferred through a dedicated PNI.

\textbf{Handover} is any AS that ingresses traffic into the eyeball network via a direct interconnection.
We denote handovers as $H_1,H_2,\allowbreak\ldots,H_j,\ldots,H_m$, where the index $j$ identifies the handover AS. 
Traffic emanating from a source AS may enter the eyeball network through one (or more) other AS(es). 
The last AS transferring traffic is the handover AS.
In this study, we focus on handover ASes that deliver traffic emanating from hyper-giant source ASes.

Traffic emanating from a source AS $S_i$ and entering the eyeball network through a handover $H_j$ is denoted as $T_{S_i,H_j}$. 
Most hyper-giant source ASes establish direct connections to large eyeball networks through \textbf{private network interface (PNI)} links as the dominant traffic delivery channel.
In such cases the source AS $i$ is also the handover AS $i$. 
Traffic delivered through a PNI link is, therefore, denoted as $T_{S_i,H_i}$. 
We form traffic volume time-series by aggregating traffic volumes for each pair of source and handover ASes on a hourly basis. 
$T_{S_i,H_j}[t]$ denotes the total traffic volume emanating from $S_i$, entering the eyeball network under investigation through $H_j$ during the hour $t$. 
$T_{S_i,H_j}[t_1:t_2]$ denotes the time-series of hourly traffic volumes from $t_1$ to $t_2$.   


The vast majority of the traffic emanating from a hyper-giant source AS $S_i$ enters the eyeball network through its PNI: $\avg_t~T_{S_i,H_i}[t] \gg\avg_t~T_{S_i,H_j}[t]$ for any $j\neq i$.   
We refer to $T_{S_i,H_i}$ as \textbf{regular traffic} denoted by $R$ in Fig.~\ref{fig:schematic-overview-2}.
Nevertheless, a small fraction of traffic emanating from hyper-giant source ASes is constantly delivered through non-PNI links. 
The reasons for the traffic flow through non-PNI links may vary. 
For example, a CDN may incorrectly map some of the users to a distant cluster~\cite{poese2010improving,pujol2019}.   
Such mapping issues may occur, for example, if the end-user uses a web proxy while the CDN employs IP based localization or if the user default domain name system (DNS) server settings are misconfigured and the CDN employs DNS based localization.  
In addition to small scale traffic flows constantly delivered through non-PNI links, occasional flash crowd events such as massive software updates or new TV series may cause PNI link over-utilization~\cite{blendin2018}. 
The excess traffic is diverted by the hyper-giant to public peering interconnects.  
We refer to traffic $T_{S_i,H_j},i\neq j$ that should have been delivered through a PNI link but is delivered through a non-PNI link as \textbf{overflow traffic}, denoted by $O_{S_{i}}$ or $T_{S_i,H_j},i\neq j$ in Fig.~\ref{fig:schematic-overview-2}.

\begin{figure}
\centering
\includegraphics[width=7cm]{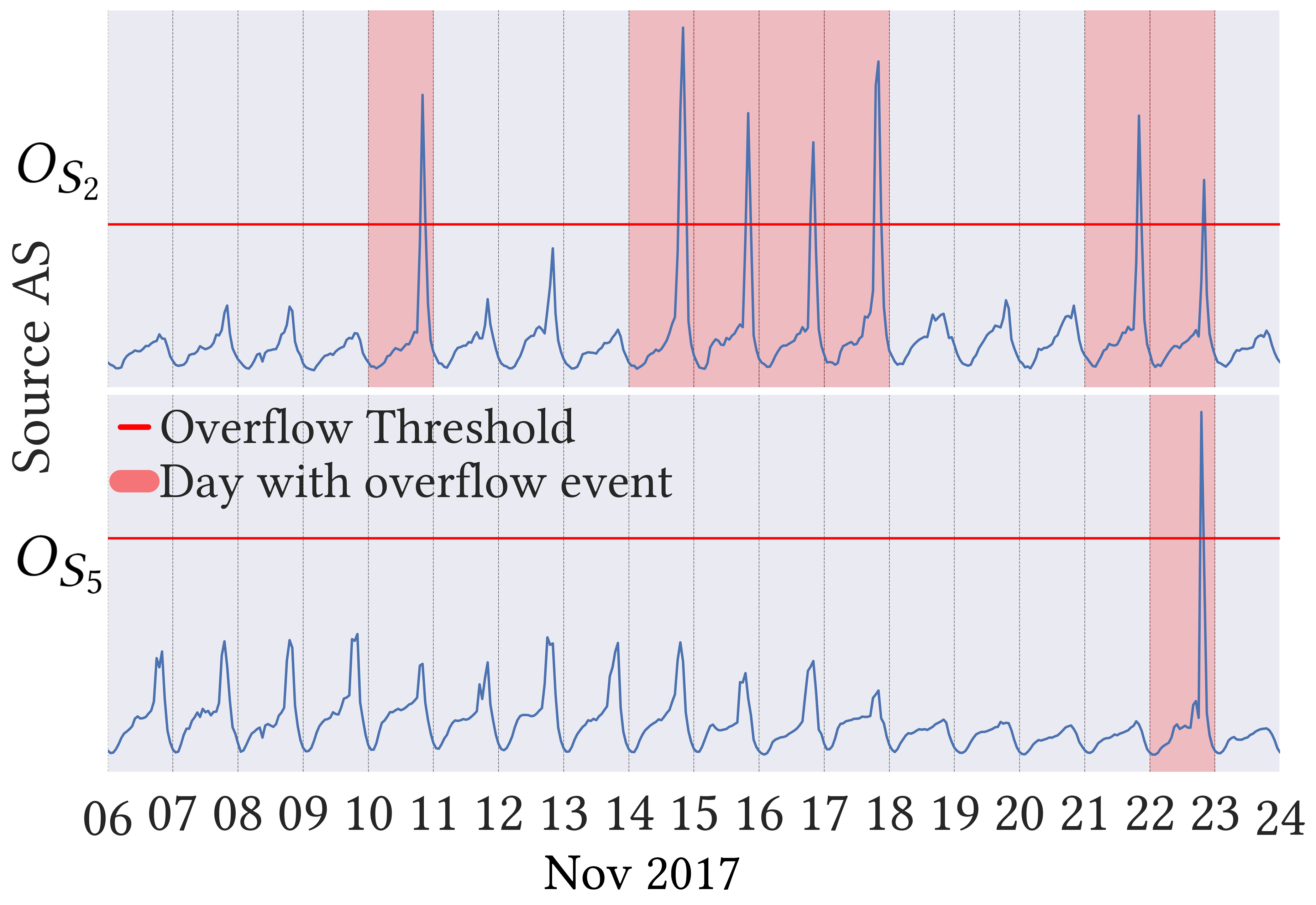}
\caption{Two exemplary overflow traffic time-series $O_{S_{2}}$ and $O_{S_{5}}$ with threshold-crossing \textit{overflow events}.}
\label{fig:overflow_illustration}
\vspace{-0.5cm}
\end{figure}

This study focuses on the rare events when the PNI links fail or their capacity is exhausted causing massive utilization of non-PNI public peering links. 
We refer to such events, depicted in Fig.~\ref{fig:overflow_illustration} as \textbf{overflow events}. 
We identify overflow events as a point in time when the total overflow traffic emanating from a source AS exceeds a certain predefined threshold $\theta_{S_{i}}$. 
It means that overflow traffic volume is abnormally high and it is most likely to cause excessive costs to the CDN and user experience degradation.
We label an overflow traffic time-series measurement at a point in time as ``overflow'' if $O_{S_{i}}[t] > \theta_{S_{i}}$ otherwise we label it as ``non-overflow''.

\textbf{No inter-hyper-giant interconnections:} While technically feasible, interconnections between hyper-giants - including the use of another CDN as a Handover AS to serve one's own traffic to the eyeball network - are \textit{not} foreseen.
Reasons for this follow business, strategic, and economic considerations, as hyper-giants intend to serve their user base located in the eyeball networks \textit{directly} via PNIs, which they have the most control over (in both technical and economic sense).
Figure \ref{fig:data_heatmap} depicts this non-occurrence of inter-hyper-giant traffic, based on our ISP data set.
This does not mean that these interconnections do not occur in the Internet.
However, when they do it is mostly due to operational fallacies or malicious behavior, and not out of business intent \cite{streibelt2018}.

\vspace{-0.2cm}
\section{Problem Definition}
\label{sec:problem_definition}
We focus on the top-10 source ASes (volume-wise) according to the selection criteria defined in the previous section, together providing more than 95\% of traffic delivered into the eyeball network, denoted as $S_1,...,S_{10}$. The top 10 source ASes employ 22 major handover ASes denoted as $H_1,...,H_{22}$. The PNI handover ASes are $H_1,...,H_{10}$ and thus, $S_i=H_i$ for $i = 1,...,10$. $H_{11},...,H_{22}$ represent the non-PNI handover ASes.
All predictions in our study are made for daily \textit{peak hours}, which we constrain to hours 17:00 - 21:00 UTC.

We define four overflow prediction problems organized along the following two dimensions: 
\begin{compactitem}
    \item Problem type: 
    \begin{compactitem}
        \item Regression problem: Predicting the volume of the overflow traffic. 
        \item Classification problem: Predicting whether an overflow event will happen, per day. 
    \end{compactitem}
    \item Data granularity:
    \begin{compactitem}
        \item Total overflow traffic: Whether the predicted variable is derived from the total overflow traffic for source AS $S_i$, i.e.  $O_{S_i}$. 
        \item Per-handover overflow traffic:  Whether the predicted variable is derived from the overflow traffic delivered through a specific handover AS $H_j$, i.e. $T_{S_i,H_j}, j \neq i$.
    \end{compactitem}
\end{compactitem}
The combination of these two factors, which define the four problems, are shown in Table \ref{tab:problems_table}. We will reference the different problems by the numbering defined in this table.

\begin{table}
  \caption{The four problems presented in the paper, regarding a certain source AS $S_i$.}
  \label{tab:problems_table}
  \begin{tabular}{p{4cm} p{4cm} p{4cm}}
   \toprule
    & \multicolumn{2}{c}{\textbf{Problem type}} \\
    \cline{2-3}\noalign{\smallskip}
  \parbox[t]{4cm}{\textbf{Granularity}} & Regression & Classification\\
    \midrule
    \parbox[t]{4cm}{Total overflow traffic $O_{S_{i}}$} 
    & \parbox[t]{4cm}{\textbf{(1)} Predict $O_{S_i}$, total overflow traffic volume} 
    & \parbox[t]{4cm}{\textbf{(3)} Predict overflow events for $O_{S_i}$}
    \\ \noalign{\smallskip}
\midrule
    \parbox[t]{4cm}{Per-handover overflow traffic $T_{S_{i},H_{j}}, j \neq i$} 
    & \parbox[t]{4cm}{\textbf{(2)} Predict $T_{S_i,H_j}, i\neq j$ per handover}
    & \parbox[t]{4cm}{\textbf{(4)} Predict overflow events for $T_{S_{i},H_{j}}, j \neq i$}
    \\
    \bottomrule
\end{tabular}
\vspace{-0.5cm}
\end{table}

\vspace{-0.2cm}
\subsection{Defining the regression problems}
The regression problems (1 \& 2) can be defined as follows:
For source AS $S_i$, at time \(t\), given $T_{S_i,H_j}[t-n:t]$ for each $H_j$ employed by $S_i$ (hourly traffic volumes from the last \(n\) hours), predict:
\begin{compactitem}
    \item Sum prediction (problem \textbf{1}): $O_{S_i}[t+h]$, the sum of overflow traffic for all non-PNI handovers employed by source AS $S_i$, at time \(t+h\).
    \item Per handover prediction (problem \textbf{2}): $T_{S_i,H_j}[t+h], i\neq j$, the overflow traffic for a certain non-PNI handover employed by source AS $i$, at time $t+h$.
\end{compactitem}
In our study, we chose \(t\)=15:00 and \(h=2,3,4,5,6\), which means we make daily predictions at 15:00 o'clock regarding the \textit{peak hours} which are 17:00-21:00. 

\vspace{-0.2cm}
\subsection{Defining the classification problems}
In the classification approach (problems 3 \& 4), each day in each overflow traffic series is labeled as either overflow or non-overflow. 
This labeling requires: 
\begin{compactitem}
    \item Defining a threshold $\theta$ for each overflow traffic time-series, above which traffic is regarded as overflowing.
    \item Taking the maximum daily \textit{peak hour} traffic amount of a total overflow traffic time-series, $O_{S_i}$ (problem \textbf{3}), or
    $T_{S_i,H_j}, i \ne j$ (problem \textbf{4}). If this daily value is above the corresponding threshold ($O_{S_i} > \theta _{S_i}$ for problem \textbf{3}, or $T_{S_i,H_j} > \theta _{S_i,H_j}$ for problem \textbf{4}), the corresponding day is labeled as ``overflow''. Otherwise, it is labeled as ``non-overflow''. The days for each overflow traffic time-series are labeled independently.
\end{compactitem}

The threshold $\theta _{S_i}$ ($\theta _{S_i,H_j}$) for each traffic volume time-series $O_{S_i}$ ($T_{S_{i},H_{j}}$) is assumed to be known in advance and to represent rare events. 
The threshold for each traffic volume time-series conforms to the formula $\theta _i = \avg~O_{S_i} + p\times \std~O_{S_i}$ ($\theta _{S_i,H_j} = \avg~T_{S_i,H_j} + p\times \std~T_{S_i,H_j}$). 
$p$ is automatically chosen for each traffic volume time-series such that at least 5\% and at most 15\% of days in the data for the overflow time-series are labeled as ``overflow''.

\vspace{-0.2cm}
\section{Proposed solution}
\label{sec:solution}
Out of the four mentioned problems, the most complex is problem \textbf{2}, where traffic volumes are to be predicted for several traffic volume time-series individually. 
From the solution to problem \textbf{2}, the solutions for all other defined problems can be derived via reduction. 
Let $T_{S_i,H_1}[t+2:t+6],...,T_{S_i,H_j}[t+2:t+6]$, for all $H_j$ employed by $S_i$ be the solutions for problem \textbf{2}, for source AS $S_i$ for a specific day. The derivations are:
\begin{compactitem}
    \item Problem \textbf{1}: by computing\\ $O_{S_i}[t+2:t+6] = \sum_{j\neq i} T_{S_i,H_j}[t+2:t+6]$.
    \item Problem \textbf{3}: by checking if $max~O_{S_i}[t+2:t+6] > \theta_{S_i}$.
    \item Problem \textbf{4}: by checking if\\ $max~T_{S_i,H_j}[t+2:t+6] > \theta_{S_i,H_j}$.
\end{compactitem}

Empirically we have found that solving the classification problems \textbf{3} and \textbf{4} by widely accepted classification methods is less accurate than predicting the exact traffic volumes by regression and applying the reductions explained above. The average AUC (area under the receiver operating characteristic curve) attained by a direct solution to problem \textbf{3}, using a classification ensemble trained by the NAS algorithm is 0.7, while the reduction approach explained yields 0.82.
We assume regression methods are better because by reducing the problem into binary classification, the  magnitude of each overflow is disregarded and the models are less able to adapt to various overflow situations. Thus, all results shown in our study are achieved by regression, while classification measures reported are extracted using the reduction explained above.

\vspace{-0.3cm}
\subsection{Applying neural architecture search (NAS)}
We chose to apply deep learning to predict the overflow traffic volumes, after using classical machine learning methods for time-series failed to give satisfactory results. Moreover, neural networks have been shown to produce state-of-the-art results for related traffic volume prediction problems given enough data (see Related Work section). We wished to derive the neural network architectures automatically to further enhance performance, and for this reason, we chose to apply a NAS method, developed by Rapaport et al. ~\cite{rapaport2019eegnas}.

This NAS method is a genetic algorithm that searches for good convolutional neural network (CNN) architectures, given data and an optimization problem. It was found to generate high-performance models for predicting time-series data in domains including finance, health, and sensor data.

The main outline of the NAS method is:
\begin{enumerate}
    \item Generate a population of \(N\) random CNN architectures.
    \item In each generation, select the better performing architectures on a hold-out validation set, perform crossover (combining two architecture to create a new one) and mutation (performing a minor alteration to an existing architecture)
    and repeat for a fixed number of generations.
    \item The selected architecture is the best performing one in the last generation.
\end{enumerate}

The output of the NAS algorithm are CNN models, which are architectures + learned model parameters, as learned throughout the NAS algorithm. The architectures are built from the most basic building blocks used in deep learning, including convolutional layers, activation functions, batch normalization, and dropout. We improved the method by creating an ensemble of the top 5 models generated in the last generation of the genetic algorithm. This increased the method's performance, as opposed to just taking the best model. The generated architectures are in table \ref{tab:NAS_archs}.

The ``total overflow traffic'' problem required models with an output size of 5, for each hour in \textit{peak hours}. In the ``per handover overflow traffic'' problem we used an output size of 5 $\times$ 22, for each hour in \textit{peak hours} for each handover AS.

\begin{table}[]
\caption{Top 5 model architectures generated by NAS for predicting overflow traffic per handover AS. Conv = 1D convolutional layer with a stride of 1 (F = number of filters, K = kernel size), MaxPool = 1D max pooling layer (K = kernel size, S = stride), ELU is a non-linear activation function by Clevert et al. \cite{clevert2015fast}.}
\label{tab:NAS_archs}
\centerline{
\begin{tabular}{@{}llllll@{}}
\toprule
Layer & Model 1           & Model 2           & Model 3           & Model 4           & Model 5           \\ \midrule
1     & Conv F: 28,K: 12  & ELU               & ELU               & ELU               & ELU               \\
2     & ELU               & ELU               & ELU               & ELU               & ELU               \\
3     & ELU               & ELU               & ELU               & ELU               & ELU               \\
4     & MaxPool K: 1,S: 2 & MaxPool K: 1,S: 2 & MaxPool K: 1,S: 2 & MaxPool K: 1,S: 2 & MaxPool K: 1,S: 2 \\
5     & ELU               & ELU               & ELU               & ELU               & ELU               \\
6     & MaxPool K: 1,S: 3 & Conv F: 37,K: 2   & Conv F: 37,K: 2   & Conv F: 37,K: 2   & Conv F: 37,K: 2   \\
7     & ELU               & ELU               & ELU               & ELU               & ELU               \\
8     & ELU               & ELU               & MaxPool K: 2,S: 3 & ELU               & ELU               \\
9     & ELU               & MaxPool K: 2,S: 3 & ELU               & MaxPool K: 2,S: 3 & MaxPool K: 2,S: 3 \\
10    & Conv F: 110,K: 39 & Conv F: 110,K: 40 & Conv F: 110,K: 40 & Conv F: 110,K: 40 & ELU               \\
11    &                   &                   &                   &                   & Conv F: 110,K: 40 \\ \bottomrule
\end{tabular}}
\vspace{-0.3cm}
\end{table}




\vspace{-0.2cm}
\subsection{Hyper parameter tuning}
\begin{figure}
\centering
\includegraphics[width=8cm]{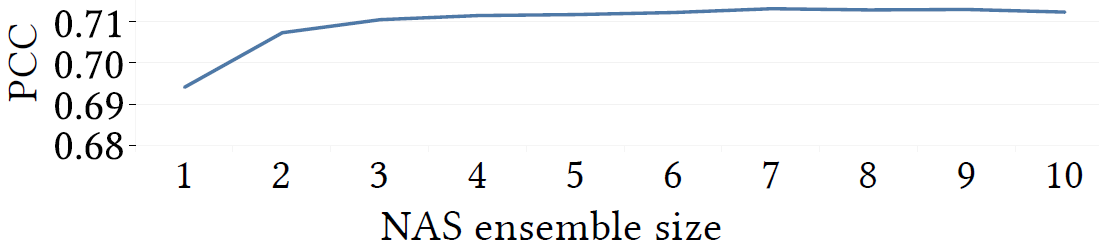}
\caption{Performance of different ensemble sizes. PCC--Pearson correlation coefficient.}
\label{fig:ensemble_sizes}
\vspace{-0.5cm}
\end{figure}

The ensemble of CNNs created by the NAS method is a simple average of model outputs. An additional weighted ensembling method was tested, using a fully connected layer which receives the outputs of all models and produces a final weighted output, but it underperformed the simple average. Ensemble sizes of 1 to 10 models were tested, and a steady increase in prediction performance was observed as the ensemble size increased. The performance improvement was very minor when using over 5 models, as shown in figure \ref{fig:ensemble_sizes}, hence this number was chosen. 

All models in the ensemble were trained for 50 epochs of the data, with early stopping after 3 non-improving epochs. The Adam \cite{duchi2011adaptive} optimizer and mean squared error (MSE) loss are used for neural network training. A window size of 240 hours for the input data was chosen empirically after testing the performance of 120, 240, and 480-hour windows. Thus, each input sample representing data delivered from a certain source AS is a matrix of size $240 \times 22$.
All code was written in Python 3.7, using the PyTorch deep learning framework. 

\vspace{-0.2cm}

\section{Evaluation and results}
\subsection{Data description}
We use hourly aggregated \textbf{flow-based data}, which is data regarding IP network traffic flows, as they enter and exit network interfaces at the edge of the AS .
The data was gathered inside a Tier-1 ISP over 2.5 years by the \textit{Flow Director (FD)} system \cite{pujol2019}.
\textit{FD} collects multiple flow-based protocols (sFlow, NetFlow and IPFIX) and produces a harmonized, well-formatted, de-duplicated, in-order flow stream that is the data basis for the hourly aggregated time series we use in this study.
The sample rate configured on router interfaces was equal on all interfaces in the ISP's AS at a rate of one to a low 4-digit number of packets.
The users requesting content via the ISP network are all located in the same time zone.
For further details on the various steps of the data processing pipeline of \textit{FD} we refer to \cite{pujol2019}. 

Each data sample given to the neural network ensemble is a 2D matrix representing 240 hourly measurements of traffic volume for each handover AS employed by a certain source AS.
Each source AS employs a specific set of handover ASes, thus a common data representation is needed in order to train the models with multiple ASes simultaneously. 
We defined each data sample as a matrix of size \(240 \times 22\). 240 represents time, for 240 hourly samples, and 22 for all of the handover ASes (240 rows and 22 columns). Only the columns representing handover ASes which are employed by a certain source AS contain real data, while other, irrelevant, rows are filled with zeros.
Formally, in all samples representing source AS $i$, the columns representing $T_{S_i,H_i}$ and $T_{S_i,H_j}, i \ne j$ for all handover ASes $j$ employed by source AS $i$ contain real data, while other columns contain zeros



\begin{figure}[ht!]
\centering
\includegraphics[width=12cm]{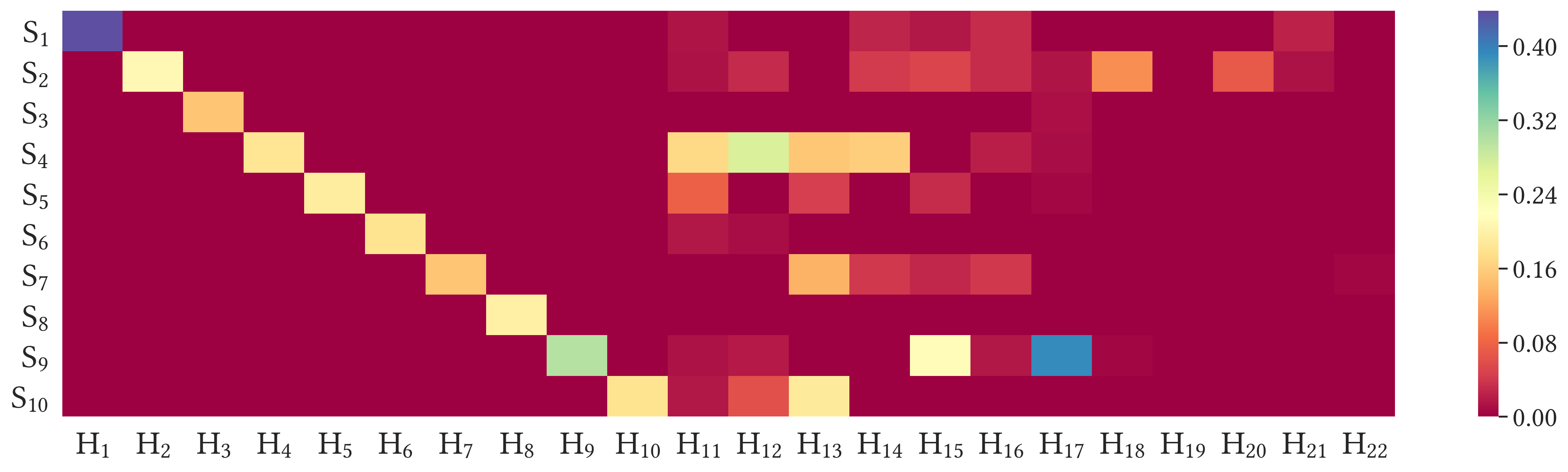}
\caption{Shown is the normalized mean traffic flow for all source/handover AS combinations, over the 2.5 years of data in the study. The diagonal in the first ten columns shows the PNI handover AS of each source AS, from which the majority of data for each source AS is delivered. The latter columns show the non-PNI handover ASes which are employed by each AS. 
}
\label{fig:data_heatmap}
\vspace{-0.5cm}
\end{figure}

By using the above data representation, the weights of the neural networks capture the patterns unique to each handover AS, which might be common across several source ASes that employ the same handover AS. Figure~\ref{fig:data_heatmap} depicts the handovers used by each AS and the intensity of the data in each AS/handover combination (after normalization).

In order to reduce the effect of the large volume of non-overflow PNI traffic volume on the prediction, the data for each traffic volume series $T$ was normalized prior to prediction. The normalization was done as follows:\\
$T_{normalized} = (T - min~T) / (max~T - min~T)$.\\
With $min~T$ ($max~T$) being the minimum (maximum) hourly measured volume for $T$ in the model training data. The ensemble received data from $T_{normalized}$ as input.

\vspace{-0.2cm}
\subsection{Experiments}
The experiments in this study are intended to evaluate the traffic prediction capabilities of the deep learning ensemble. They were conducted using 10-fold cross-validation in time, using forward chaining, similar to the methods shown by Bergmeir et al \cite{bergmeir2012use}. This means that each traffic volume time-series is split into 11 chronological parts indexed by [1, 2, 3, 4, 5, ..., 11]. We train 10 models:
\begin{compactitem}
    \item Model 1: Train data: \([1]\), Test data: \([2]\)
    \item Model 2: Train data: \([1, 2]\), Test data: \([3]\)
    \item Model 3: Train data: \([1, 2, 3]\), Test data: \([4]\), and so on...
\end{compactitem}
The reported results were calculated by combining all test data predictions into one timeline, thus receiving predicted data for segments [2, 3, 4, ..., 11], for all source ASes, and comparing them with the actual traffic volumes for the time period represented by these data. This evaluation approach ensures that our prediction model is trained using only past data, as in real-life situations. Additionally, a unique model was developed by the NAS algorithm for each fold of the data, thus the architectures of the neural networks are not affected by future data. Except for figure \ref{fig:results_per_hour}, all results depict the average predictive performance for all of the \textit{peak hours} (17:00-21:00). Furthermore, all results are based on cross-validation in time, using the whole 2.5 years of available data.

\vspace{-0.2cm}
\subsection{Traffic volume prediction by regression}
\begin{figure}[ht!]
\centering
\includegraphics[width=1\textwidth]{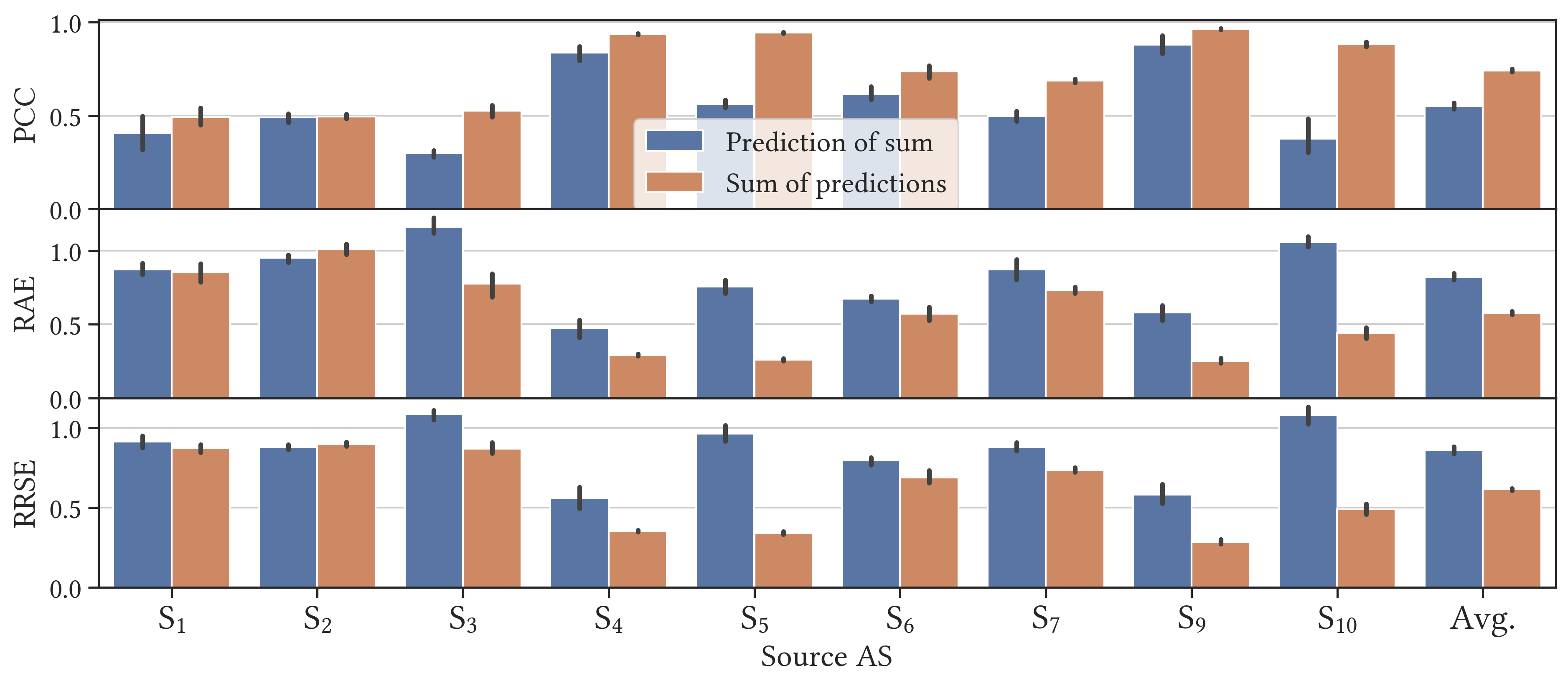}
\caption{Pearson correlation coefficient (PCC), relative absolute error (RAE), and root relative squared error (RRSE) for the prediction of traffic overflow volumes in the 10 source ASes in our study. Results are averaged for the combined 10-fold test set in 5 independent ensemble training runs.}
\label{fig:regression_results_fig}
\end{figure}

To evaluate our deep learning ensemble predictor for overflow traffic we used the following measures, which are widely accepted in the domain of MTS forecasting.
\begin{equation}
\begin{aligned}
    \text{Root relative squared error (RRSE)} = \frac{\sqrt{\sum_{(t) \in \Omega_{Test}}(T[t] - \hat{T[t]})^2}}{\sqrt{\sum_{(t) \in \Omega_{Test}} (T[t] - \bar{T})^2}}
\end{aligned}
\label{eq:RRSE}
\end{equation}
\begin{equation}
\begin{aligned}
    \text{Root absolute error (RAE)} = \frac{\sum_{(t) \in \Omega_{Test}}|(T[t] - \hat{T[t]})|}{\sum_{(t) \in \Omega_{Test}} |T[t] - \bar{T})|}
\end{aligned}
\label{eq:RAE}
\end{equation}
Pearson correlation coefficient (PCC) = \\
\begin{equation}
\begin{aligned}
  \frac{\sum_{(t) \in \Omega_{Test}}(T[t]-\bar{T})(\hat{T[t]}-\bar{\hat{T}})}{%
        \sqrt{\sum_{(t) \in \Omega_{Test}}(T[t]-\bar{T})^2}\sqrt{\sum_{(t) \in \Omega_{Test}}(\hat{T[t]}-\hat{\bar{T}})^2}}
\end{aligned}
\label{eq:RAE}
\end{equation}
where $T$ denotes a traffic volume time-series, $\hat{T}$ denotes predicted volumes for $T$, and $\bar{T}$ denotes the average \textit{peak hour} values for $T$ in $\Omega_{Test}$, the total evaluated time period. The above measures are taken for the daily predicted \textit{peak hours} - daily measurements for the hours 17:00 to 21:00.

\begin{figure}
\centering
\includegraphics[width=1\textwidth]{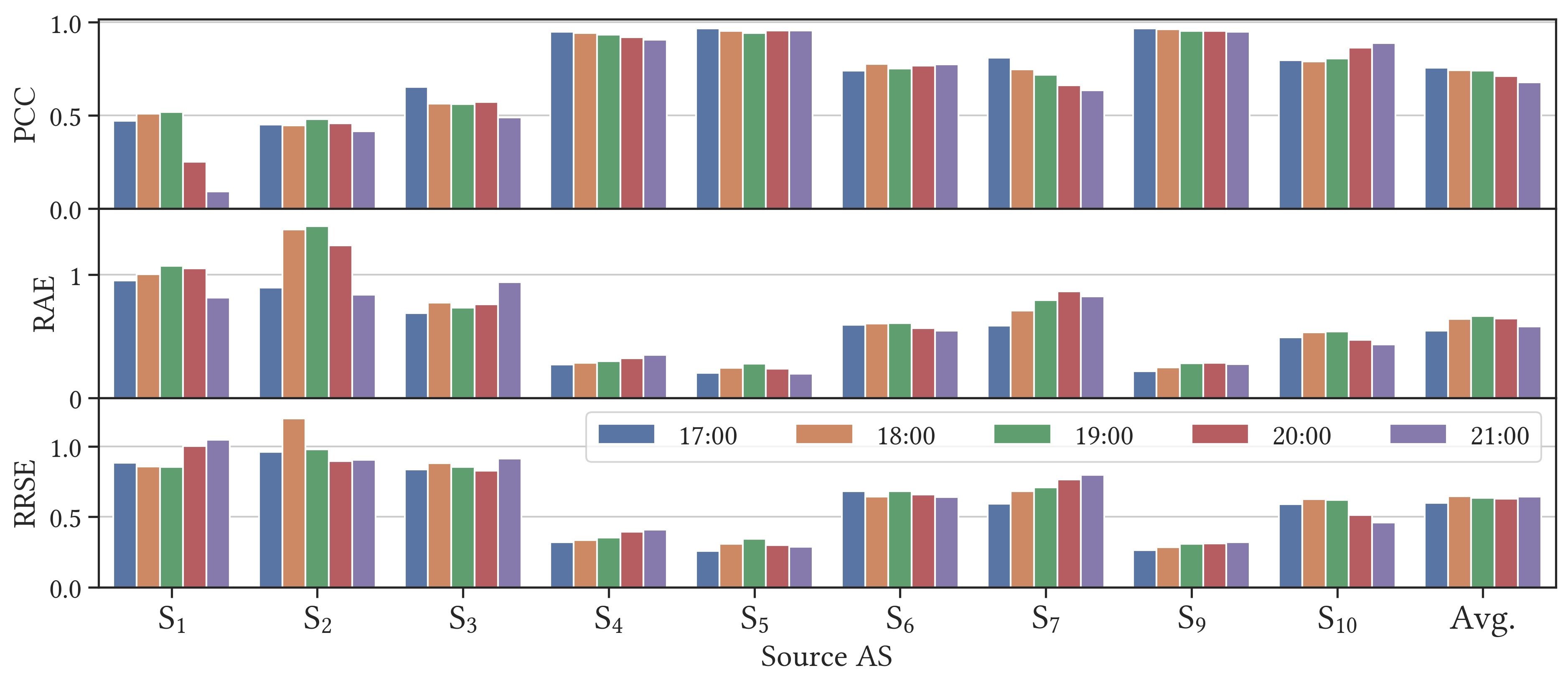}
\caption{Regression results per hour of day. It can be seen that, in average, prediction for the earlier hours is more effective.}
\label{fig:results_per_hour}
\end{figure}

Results for problem \textbf{1} (predicting the volume of ``total handover traffic'') in the 10-fold experiment are shown in figure \ref{fig:regression_results_fig}. Compared are two approaches to solving this problem. The first is by directly predicting the sum of all non-PNI handover traffic--prediction of sum, and the second is making predictions for each non-PNI handover individually and summing the results--sum of predictions. Source AS $S_8$ does not appear in our results because it does not employ any non-PNI handover ASes and thus generates no overflow traffic.

Regression results per hour of day are shown in figure \ref{fig:results_per_hour}. Prediction for the earlier hours is generally more effective than for later hours in the day. These results are sensible because the earlier hours in the \textit{peak hours} are closer to the time window of the training data.

It is shown that summing the predictions for all overflow traffic volume series ($T_{S_i, H_j}, i \ne j$) is better than predicting their sum ($O_{S_i}$). We assumed this is because when predicting the values for each overflow traffic volume series individually, the models can account for the inter-dependencies between the times series of different handover ASes, by learning the convolutional weights corresponding to different handover ASes simultaneously. When we aggregate the the time series of different ASes into a sum ($O_{S_i}$) before model training, the models lose the capability to account for inter-dependencies between ASes.

\begin{figure*}
\centering
\makebox[\textwidth][c]{\includegraphics[width=1\textwidth]{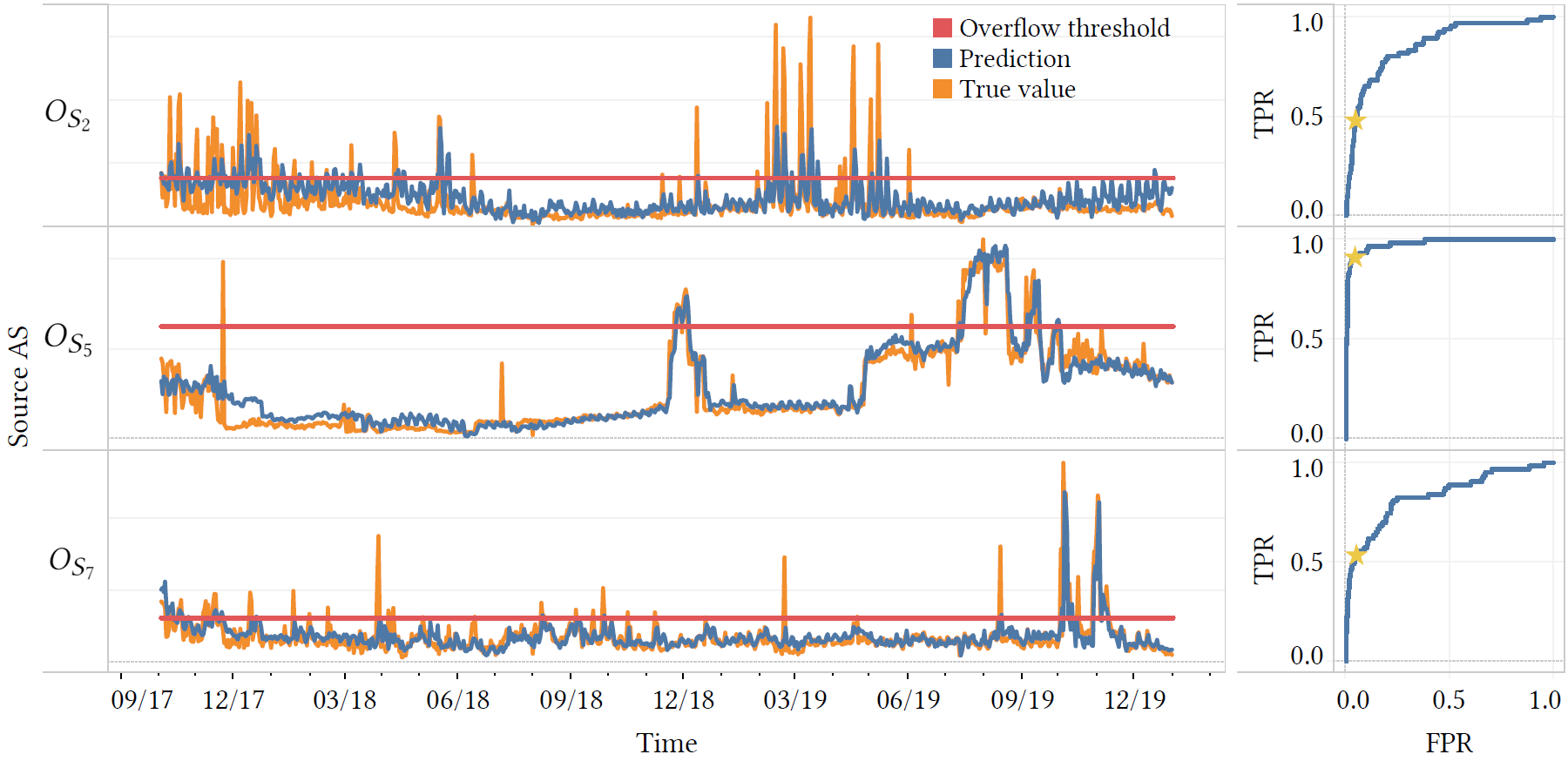}}%
\caption{Left: Overflow traffic predictions for 3 source ASes (for problem \textbf{1}). Right: The AUC-ROC curve regarding the threshold value $\theta _i$ shown on the left. Starred is the optimal point for an overflow prediction system (problem \textbf{3}), with minimal false-positives (<0.5\%). Traffic volume units were omitted due to privacy constraints. Time is in mm/yy.}
\label{fig:regression_results}
\end{figure*}

Prediction results along time are shown graphically for a selected 3 ASes in figure \ref{fig:regression_results}. Shown are the results for the sum of predictions, for the concatenated test sets taken from 10-fold cross-validation in time.

\begin{figure}
\centering
\includegraphics[width=1\textwidth]{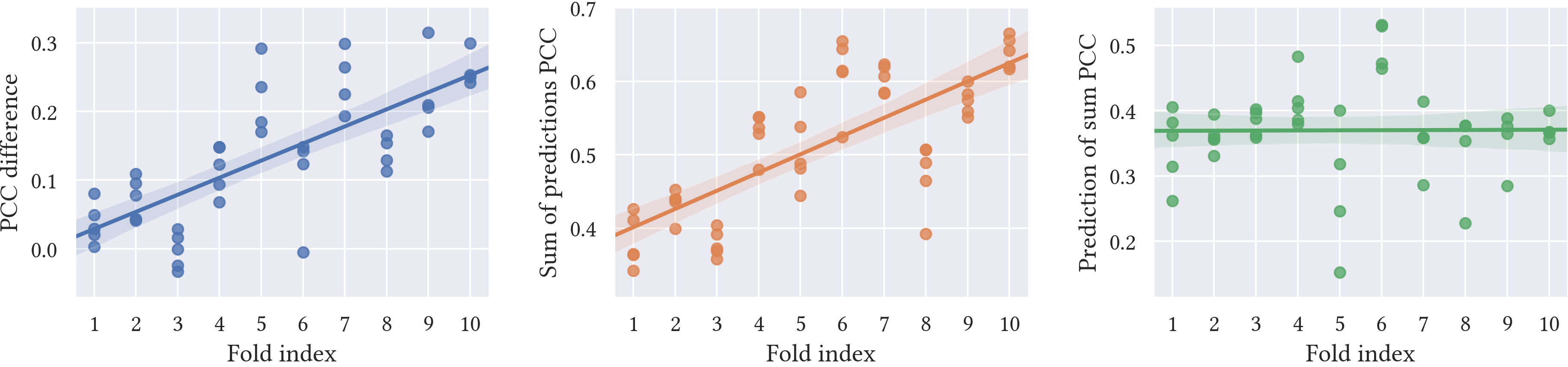}
\caption{(Right + middle): The performance of both prediction methods with regard to the fold index, which represents the amount of training data available to the network. (Left): The difference in Pearson correlation results for overflow traffic prediction between the sum of predictions and prediction of sum, by data fold.}
\label{fig:corr_diff}
\end{figure}

The notion that predicting the overflow traffic for each traffic volume series helps model the dependencies between them is strengthened in figure \ref{fig:corr_diff}. In this figure are results for 5 independent experiment runs for each fold, and the linear trend observed. It is visible that the difference between prediction of sum and sum of predictions increases in the higher folds of the 10-fold experiment.

In the higher folds more training data becomes available to the models (higher folds contain more training data). We assume that as more training data became available, the neural networks were able to better make use of the inter-dependencies between the output variables by incorporating this knowledge into the convolution weights. The sum of predictions approach improves drastically as more data becomes available, while the prediction of sums approach achieves similar performance for all training set sizes.

\begin{figure}[ht!]
\centering
\includegraphics[width=12cm]{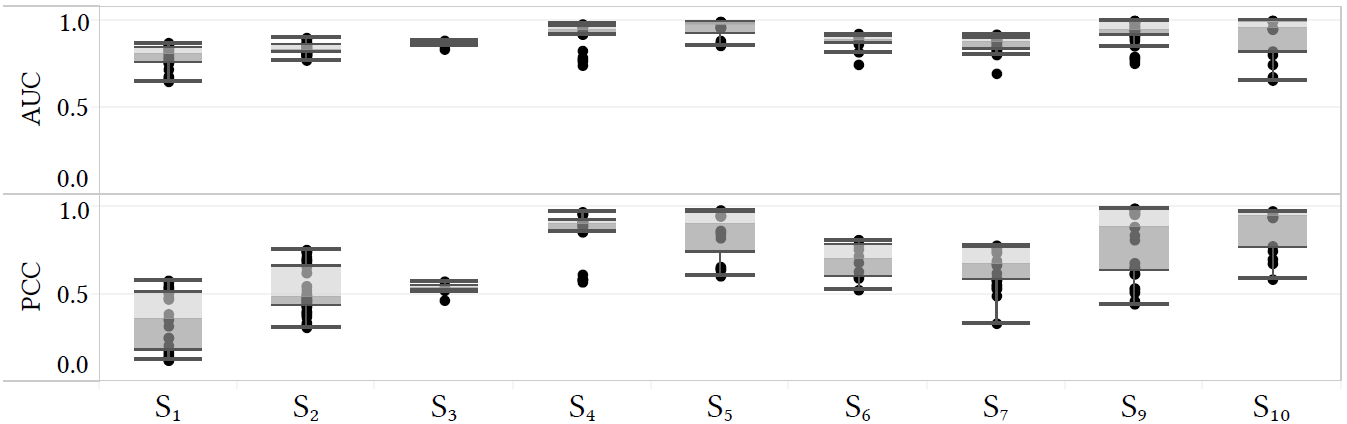}
\caption{PCC and AUC results for the different source ASes. The points in a certain $S_i$'s column represent prediction results for the different non-PNI handover ASes employed by $S_i$, in each of 5 experiment iterations.}
\label{fig:per_handover_corr}
\vspace{-0.2cm}
\end{figure}

In addition to the advantage in prediction accuracy for problem \textbf{1}, per-handover resolution prediction gives higher quality insights for the final goal of mitigating overflow events, thus solving problems \textbf{2} and \textbf{4}.
Shown in figure~\ref{fig:per_handover_corr} are detailed results for these problems for each source AS. 


The caveat in measuring only the performance of the regression task is that the accepted measures might miss the essence of the overflow prediction problem. In figure~\ref{fig:regression_results_fig} we can see, for example, that for AS $S_2$ our method scored an average RRSE of nearly 1 for the sum of predictions task. According to the RRSE measure, our method is almost worse than the naive predictor (with RRSE = 1), which predicts the average value at all times. Despite this result, applying the threshold $\theta _{S_2}$ to the predicted values and measuring the performance by classification metrics yielded a respectable AUC of 0.85 for the same predicted values. Thus the regression targeted statistic of RRSE failed to capture the essence of the overflow prediction task and additional measures are needed.

\vspace{-0.2cm}
\subsection{Overflow event prediction by classification}
\begin{figure}[ht!]
\centering
\includegraphics[width=1\textwidth]{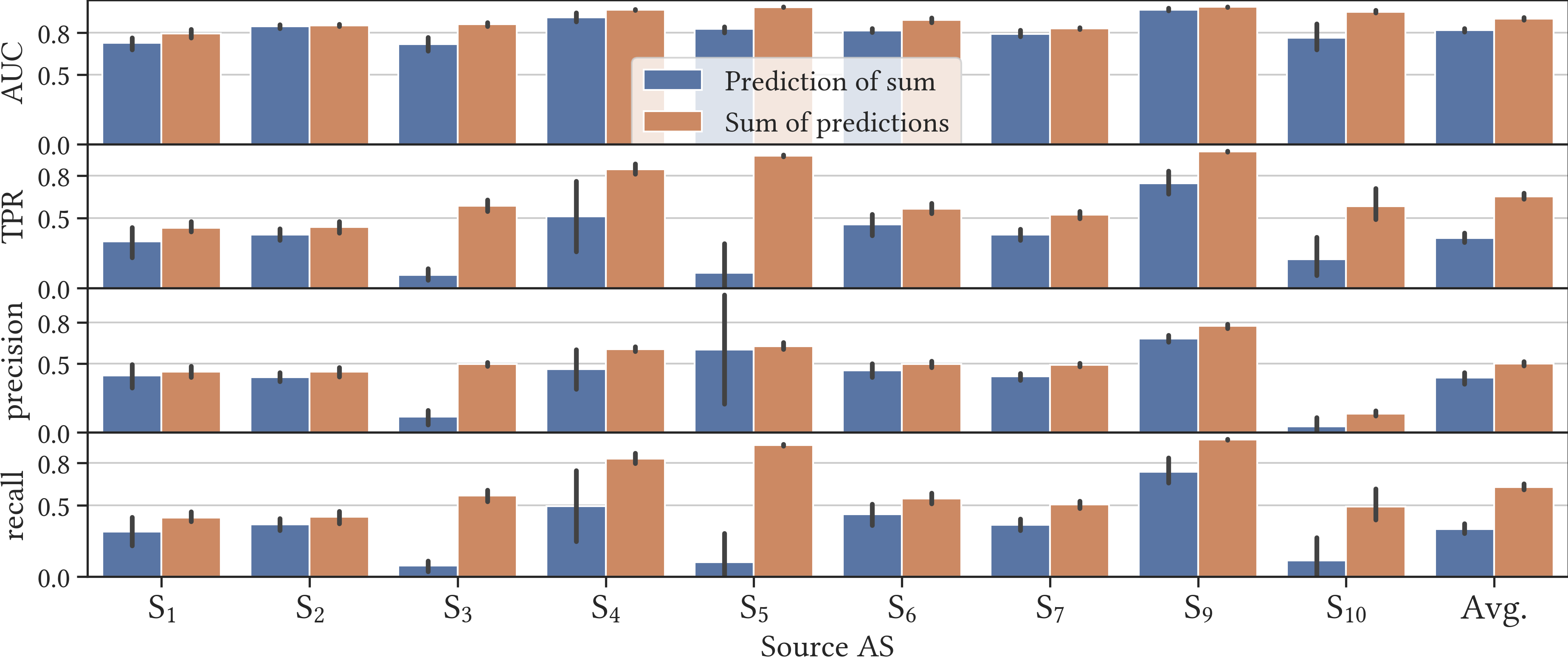}
\caption{AUC, TPR, precision and recall values for the binary classification problem of traffic overflow prediction in each source AS. Results are averaged for the combined 10-fold test set in 5 independent ensemble training runs. All TPR, precision and recall values reported relate to the point we defined as optimal, where the FPR is just below 0.05.}
\label{fig:classification_results_fig}
\vspace{-0.2cm}
\end{figure}

All classification measures were obtained by solving the regression problem and applying a threshold on the results. Shown in figure~\ref{fig:classification_results_fig} are the results for the overflow classification task, regarding $O_{S_i}$, the sum of overflow traffic in all non-PNI handovers, for each source AS $i$. Providing binary output of whether an overflow will occur for a specific day or not gives actionable insights which may be translated into the prevention of the predicted overflow.

As in the regression task, summing the predictions of each handover yielded better results than predicting the sum of overflow traffic for all handovers. Despite the low RRSE and RAE results for several source ASes, we achieved a usable true-positive rate (TPR) for almost all source ASes while maintaining a very low false-positive rate (FPR) of below 5\%.

Two different thresholds for the classification of overflow events exist. The first is the external threshold, defined by us as $\theta _i = \avg~O_{S_i} + p\times \std~O_{S_i}$ ($\theta _{S_i,H_j} = \avg~T_{S_{i},H_{j}} + p\times \std~T_{S_{i},H_{j}}$), with $p$ defined such that at least 5\% and at most 15\% of days are labeled as ``overflow'' for each overflow traffic series. The external threshold defines what is an overflow, meaning when does an abnormal peak in overflow traffic occur. The second threshold is internal to the prediction model. It is the model's decision threshold, as shown in the receiver operating characteristic (ROC) curve graphs in figure~\ref{fig:regression_results}. The internal threshold is always chosen such that a false-positive rate of no more than 5\% will be maintained, no matter what the external threshold is.

It is important to note that the choice of threshold created imbalanced data for the model to train on, thus we report a wide array of classification metrics (AUC, FPR, precision and recall) in our results to ensure proper model assessment.

\subsection{Comparison to state-of-the-art}
In order to validate our method's performance we conducted an evaluation against three other time series forecasting methods.

\begin{compactitem}
    \item LSTNet \cite{shih2019temporal}. An MTS forecasting RNN+CNN architecture, which achieved state of the art performance on several benchmark MTS datasets. This method leverages periodicity in the data via an attention mechanism, which accounts for the daily patterns in the data.
    \item a vanilla LSTM \cite{hochreiter1997long} recurrent neural network model, containing four LSTM layers with a hidden dimension of 100 and a "dense" fully connected output layer.
    \item A CNN architecture automatically found by NSGA-Net by Lu et al. \cite{lu2019nsga}, another SOTA NAS algorithm. This algorithm searches for good structures of CNN building blocks, and these building blocks are eventually combined in a predefined manner to form a final network.
\end{compactitem}

\begin{figure}[ht!]
\centering
\includegraphics[width=1\textwidth]{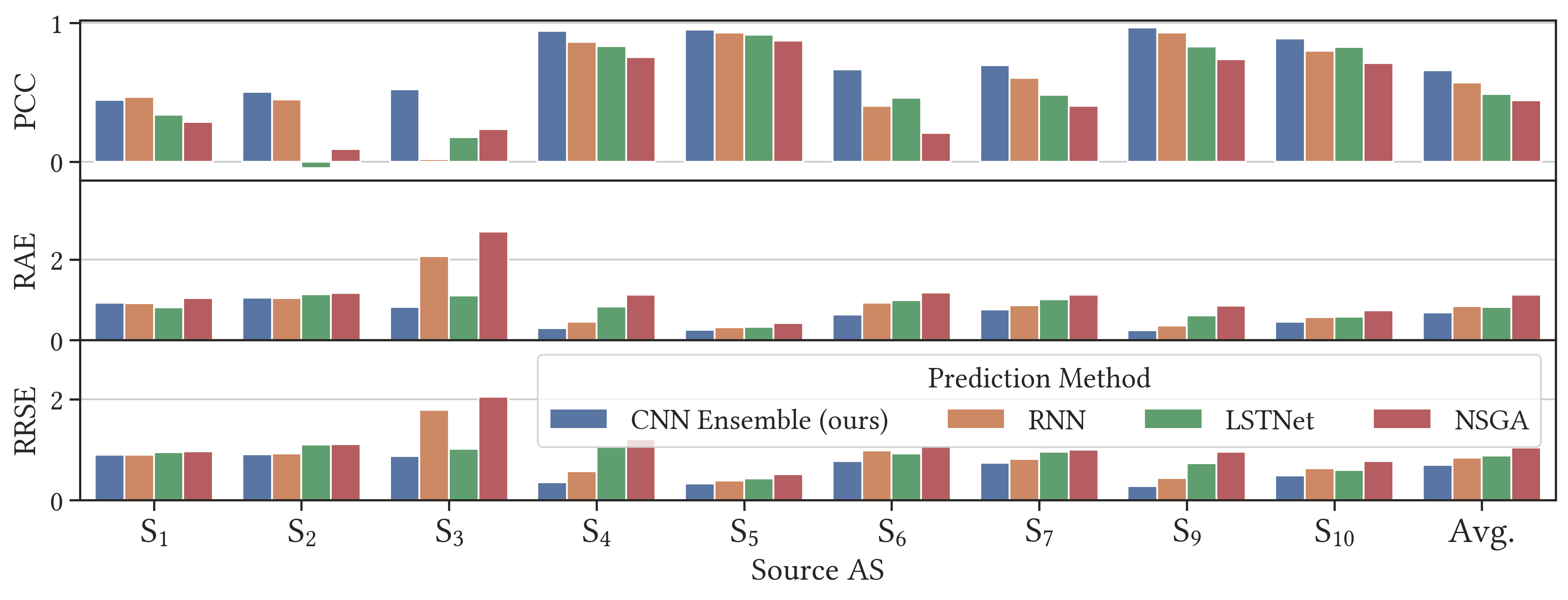}
\caption{Comparison of regression measures amongst different MTS forecasting techniques. Results are averaged for the combined 10-fold test set in 5 independent training runs per method.}
\label{fig:model_comparison}
\vspace{-0.2cm}
\end{figure}

The comparison of the different methods shown figure \ref{fig:model_comparison}  shows that ours is the most accurate.

\vspace{-0.2cm}

\subsection{Feedback from the ISP on method performance}
We have shared the method and prediction results with network operations experts at the ISP and requested feedback.
The principal capability of predicting overflow events a few hours in advance was generally perceived as a valuable means toward one of the ISP's strategic objectives: to be a market leader in customer experience regarding service usage over its network.
Customer experience is linked to overflow events because many popular Internet services including social media and online gaming are delivered via hyper-giant infrastructures.
There is a substantial risk of some traffic flowing over non-PNIs (i.e. an overflow event), possibly meeting undersized interfaces and ultimately leading to service degradation.
Very popular services can easily be consumed by 1-digit million users that are customers of the ISP -- and are potentially exposed to this risk.
Negative customer experiences today quickly surface on social media and are associated with the ISP in public, negatively impacting the ISP's brand perception.
Therefore, any measures, that accelerate the successful mitigation of such incidents, positively affect the ISP's service quality perception and minimize the the complaints.
The predictive method presented here gives a chance to avoid service degradation due to overflow events all together if the mitigation measures (section~\ref{mitigation}) are effective and fast enough.
As to the current prediction accuracy of the method the ISP network experts rated the TPR of more than 0.75 for some of the major hyper-giant ASes acceptable.
Yet, it needs to be improved over time.
The pegging of the FPR to 0.05 was highlighted as an important feature in order to avoid unnecessary alarms to the network operations crew and to increase acceptance levels regarding the method.

\subsection{Overflow prediction during the COVID-19 outbreak}
The COVID-19 pandemic has brought with it a major change in lifestyle for humans around the globe. Restrictions such as social distancing and cancellations of public gatherings, with the encouragement to stay home, had noticeable impact on network traffic, with people relying on online tools to stay connected, and consume information and entertainment.

\begin{figure}[ht!]
\centering
\includegraphics[width=12cm]{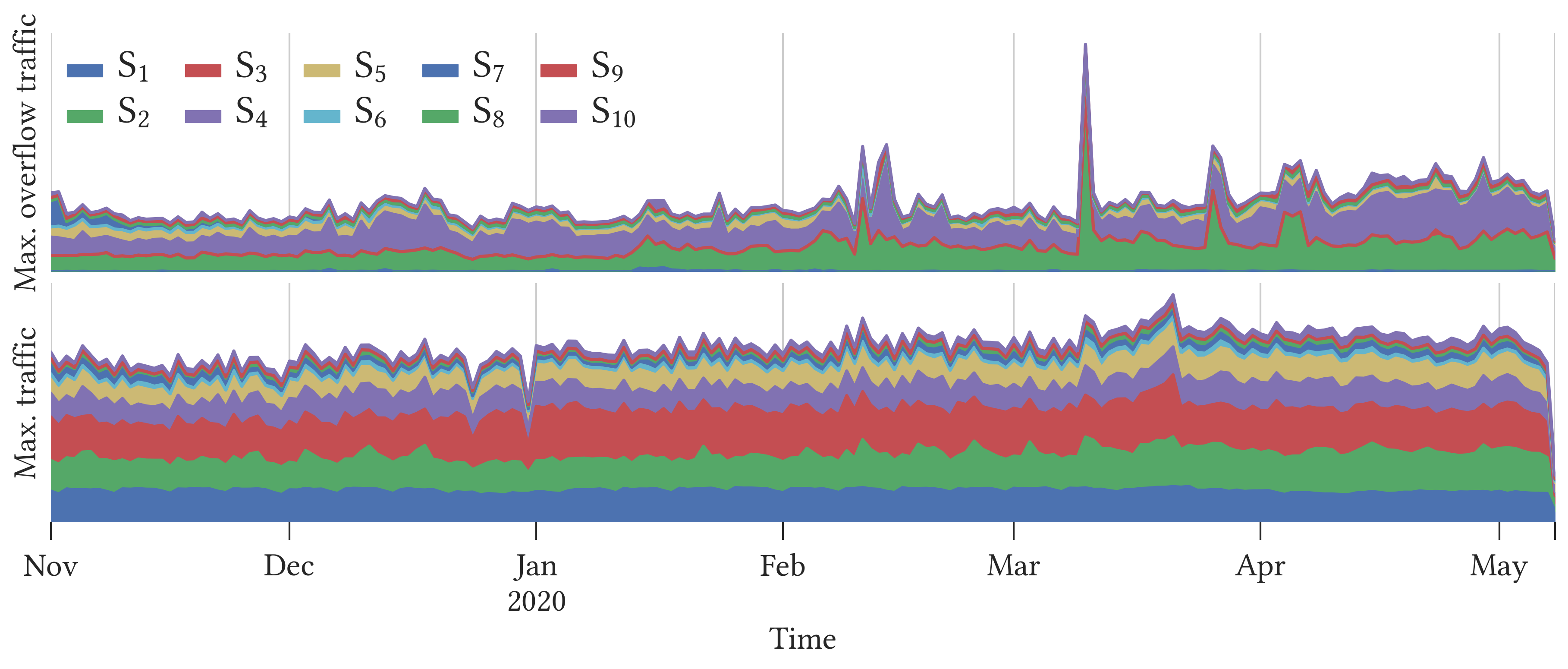}
\caption{Daily NetFlow traffic for the top-10 ASes before and during the COVID-19 crisis. Top: sum of the overflow traffic series. Bottom: Sum of all traffic volume series, including PNIs. Time is in mm/yy.}
\label{fig:corona_viz}
\vspace{-0.2cm}
\end{figure}

Figure~\ref{fig:corona_viz} (bottom) shows the total ingress traffic into the ISP from the 10 source ASes before and during the COVID-19 pandemic. A clear rise in traffic is seen in mid-March, about the time the new restrictions took place across the world.
Figure~\ref{fig:corona_viz} (top) shows the overflow traffic rising steadily in the same period, which means that in addition to heightened traffic volumes, probably more overflow events occurred during the COVID-19 outbreak.

\begin{figure}
\centering
\includegraphics[width=1\textwidth]{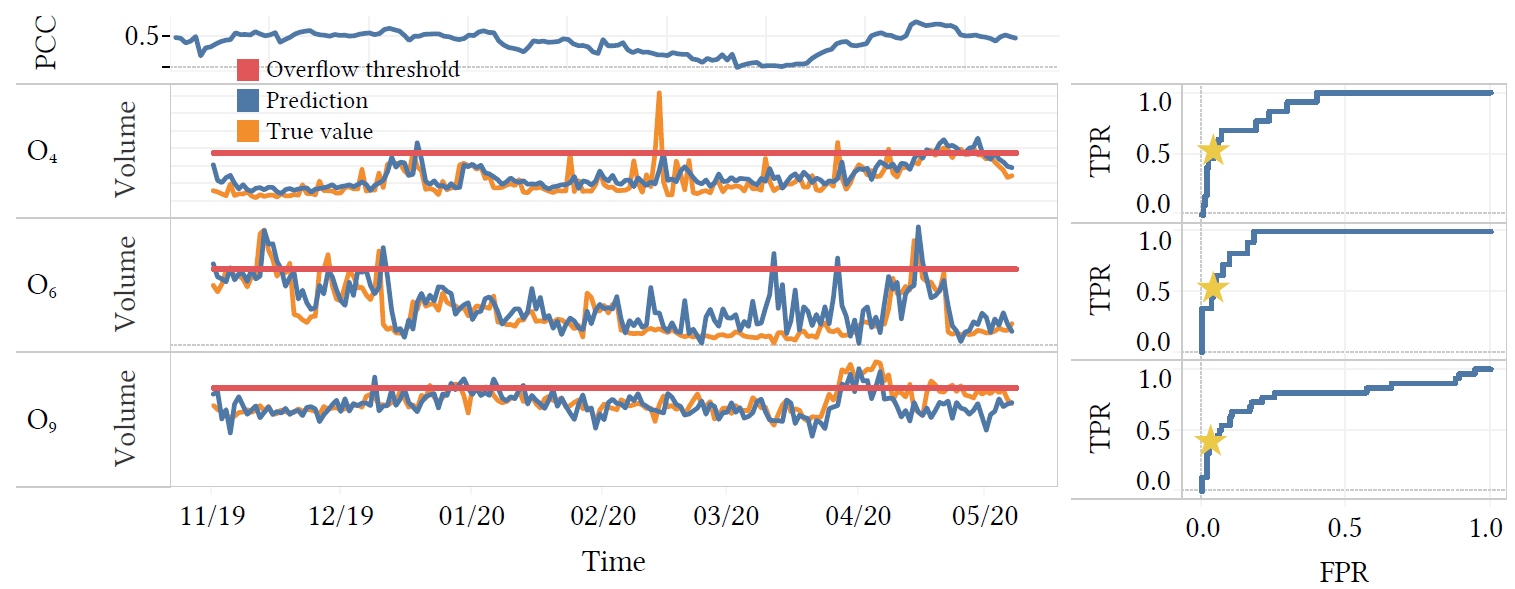}
\caption{Network traffic prediction results for the COVID-19 pandemic period for three source ASes. A slight increase in traffic volume is visible from late March, around the time various countries imposed movement restrictions on citizens. Shown on top is the average prediction quality in PCC for $O_{S_4},O_{S_6},O_{S_9}$ over time. Actual traffic volumes are omitted due to privacy constraints. Time is in mm/yy.}
\label{fig:corona_results}
\vspace{-0.5cm}
\end{figure}

Figure~\ref{fig:corona_results} shows our ensemble's performance on the COVID-19 period data for problem \textbf{1} (shown are the 3 source ASes which showed the highest rise in overflow traffic in the COVID-19 period). The ensemble was trained with all data until Nov. 1st, 2019 and tested on data from Nov. 1st, 2019 onwards. Prediction quality worsens over time and this can be attributed to two main factors:
\begin{compactitem}
    \item The time range for prediction is longer than in our previous experiments. One ensemble predicts 7 months of data (as opposed to 2.5 months ahead in the 10-fold experiment).
    \item Data in the selected source ASes, from February onwards, is noisier than previous periods, most likely because of COVID-19, and its prediction is harder.
\end{compactitem}
Despite the performance degradation during the peak of the COVID-19 crisis, a clear rise in prediction performance is seen from April, around the time which governments around the globe started lifting restrictions. The rise in performance shows that our model is robust when predicting normal data which is not affected by COVID-19 era network traffic behavior, and suffers from worse performance when dealing with data which is abnormal.

\subsection{Computational resources}
\label{comp_res}
All experiments were run on an Intel Core i7-8700 CPU @ 3.20GHz machine with 32GB RAM, utilizing a single Nvidia RTX 2080 GPU with 12GB RAM.

The ensemble, including all parameters, is 4.5 megabytes in size, which allows it to fit in almost any dedicated system. Moreover, the model's inference time for predicting 2.5 months of data is 0.05 seconds on average, which is more than enough to allow reaction in case of a predicted overflow.

A complete 10-fold experiment conducted in this study takes 36 hours to run on our system. This run-time includes the genetic NAS algorithm as well. Thus it is feasible to completely retrain our model on a monthly or even weekly basis without any specialized hardware.

\vspace{-0.2cm}
\section{Deep Learning Visualization}
There has been a recent surge in interest in the field of DL explainability, in order to unveil these seemingly black-box models and try to explain the logic behind their decisions in a humanly understandable manner. Following this trend, with the aim of understanding the decision making process of our prediction neural network ensemble, we utilized three deep learning visualization techniques to understand which parts of the input data affected the prediction and to what extent. The methods we utilized are:
\begin{compactitem}
    \item \textbf{Deeplift} \cite{shrikumar2017learning} by Shrikumar et al. attains the feature attribution of each part of the input by comparing the error for each prediction with the error of a reference input (a matrix of all zeros), using gradient descent.
    \item \textbf{GradientSHAP} \cite{lundberg2017unified} by Lundberg et al. computes an approximation of SHAP values for the input data, which are based on Shapely values from game theory and have been shown to represent importance attributions.
    \item \textbf{Deconvolution} \cite{zeiler2014visualizing} by Zeiler et al. constructs a neural network by stacking the layers of the tested neural network in reverse order and computing the gradient of the target output with respect to the input. The outputs of this model are the importance attributions.
\end{compactitem}

\begin{figure*}
\makebox[\textwidth][c]{\includegraphics[width=1\textwidth]{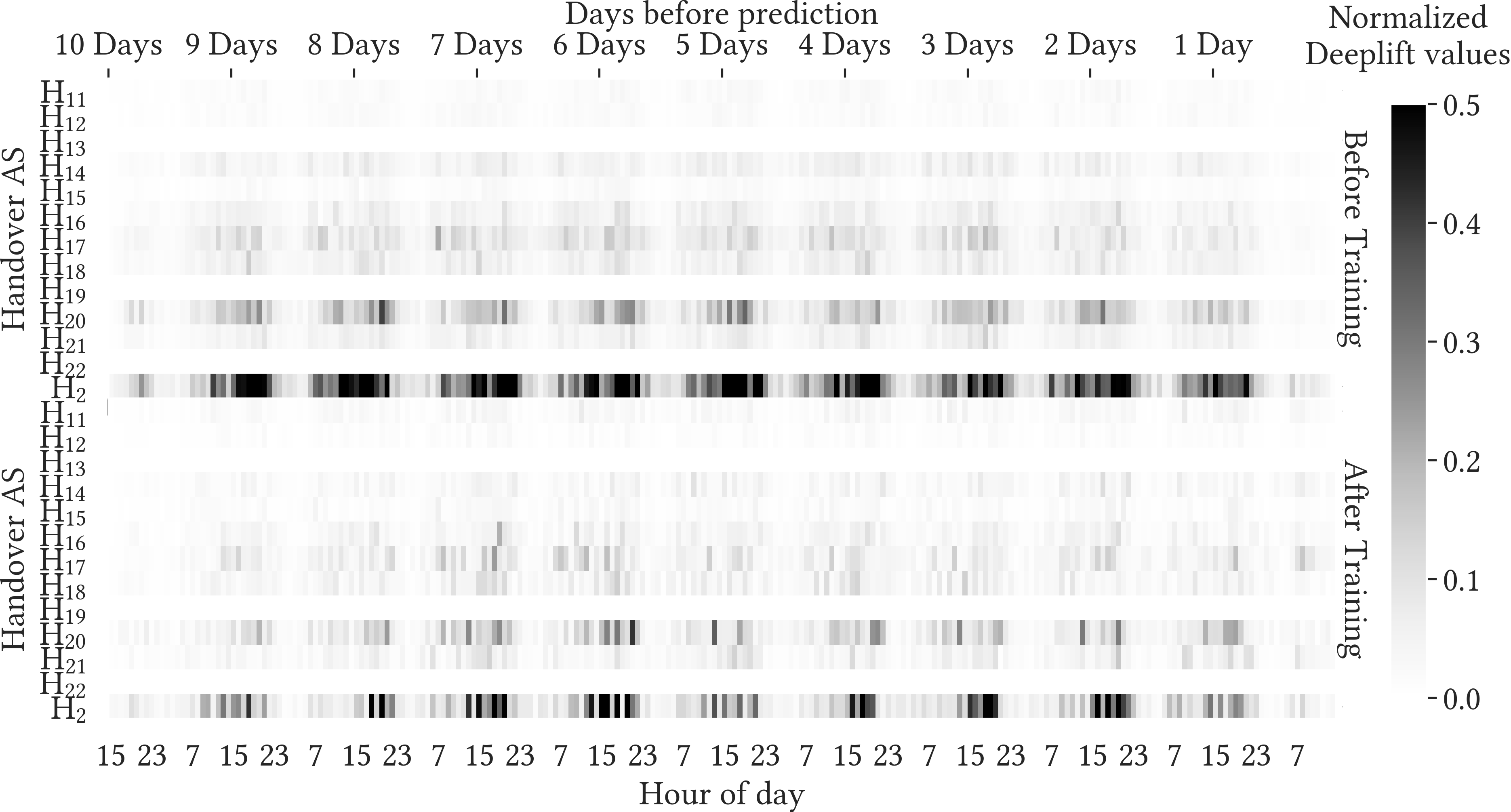}}%
\caption{Average feature importance attributions for source AS $S_2$, for all test samples of the last fold in the 10-fold experiment, calculated by the Deeplift technique. Shown are the feature importance values for a trained and untrained ensemble.} 
\label{fig:deeplift_values}
\end{figure*}

In Figure~\ref{fig:deeplift_values} we see the absolute value importance attributions for each part of the input, when using traffic volumes from $S_2$ as testing data (the ensemble was trained on all ASes), as calculated by Deeplift. The original data contains $H_1$ and $H_3,...,H_{10}$ but they were omitted from the chart since no source AS in the study employs other source ASes as handovers. The importance attribution values were calculated for every sample in the last fold of the 10-fold experiment and averaged in the visualization. The same is shown for a trained and un-trained ensemble.

The highest attribution is given to the PNI data, as seen in the last row of both parts of the figure, and this effect heightens in the trained ensemble compared to the untrained one. This means that the models are able to identify to which AS each sample belongs and there is no need to include the AS identity in the data representation.

Additionally, we can see that before training, high importance values are given to large portions of each day. After training, higher attributions are given to the predicted \textit{peak hours} of 17:00 to 21:00 in each day. It is also noticeable that the several hours prior to prediction have almost no effect in the untrained ensemble while having a reasonable effect on the trained ensemble's output. This effect coincides with the results, as shown in figure \ref{fig:results_per_hour}, where prediction performance for the hours close to the edge of the prediction window is higher. A reasonable assumption is that high performance in the earlier \textit{peak hours} of the predicted day is correlated with the high importance of observations from the predicted day.

\begin{figure}[ht!]
\centering
\includegraphics[width=1\textwidth]{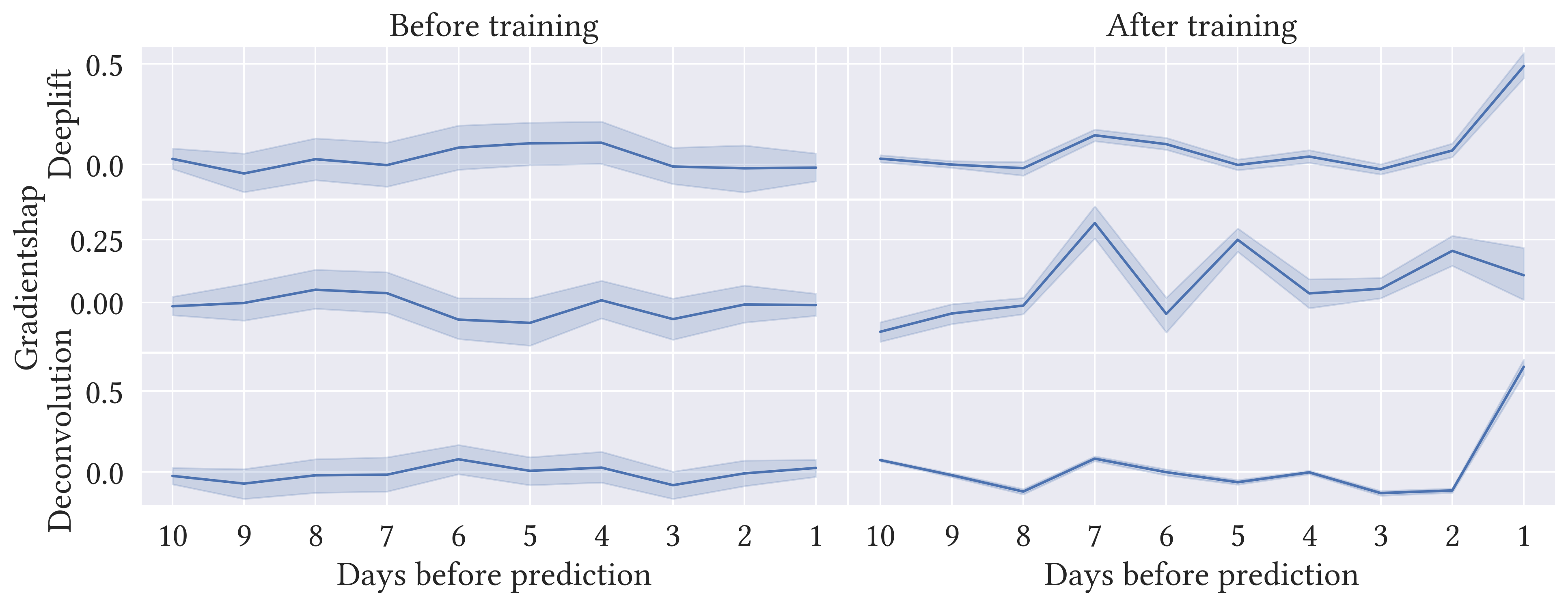}
\caption{Avg. importance attributions for: Deeplift, GradientSHAP and Deconvolution for trained and untrained ensembles, aggregated by day before prediction. Confidence intervals calculated using bootstrapping \cite{diciccio1996bootstrap}.}
\label{fig:importance_by_day}
\vspace{-0.2cm}
\end{figure}

Figure~\ref{fig:importance_by_day} shows the average importance attributions for all training samples, for the three visualization methods used, aggregated by day. Before ensemble training, all days prior to prediction were given similar importance and no clear trend is seen amongst the three methods. After training, a clear positive peak is seen a week before prediction, meaning that traffic volumes that ocurred a week prior to prediction will have a strong effect on the predicted volumes. Moreover, the confidence intervals are much smaller in the trained ensemble compared to the untrained one, which means this pattern is consistent amongst many source AS/handover AS combinations in the data.

The second visible trend (in the Deeplift and Deconvolution methods) is that traffic in the 1-2 days prior to prediction has a very high positive effect on the predicted volumes. This trend explains events such as public holidays, where video streaming traffic is high for several consecutive days.

\vspace{-0.2cm}
\section{Overflow handling}
\label{mitigation}
As overflows are induced by hyper-giants 
with enough capacity to disrupt other services, we focus our attention there. Also, handling overflows needs to be done transparently to the users of the network before performance issues arise. This means that the detection, communication, and mitigation of overflows has to be seamless in a fully automated and preemptive fashion.

In contrast, most overflows today are handled in retrospect and on the ``political'' layer of the Internet. This usually happens in one of two ways: 
\begin{enumerate}
    \item one of the involved parties notices the overflow after it has begun and contacts the other parties via human channels (i.e. email, phone) to work out the issue.
    \item a party has prior knowledge that a traffic spike is going to happen due to scheduled events (i.e. soccer games, large updates, etc) and preemptively announces that.
\end{enumerate}

Both options are reactive. In case (1) the overflow has already happened. In case (2) the overflow location is not yet known and needs to be handled once it arises, as in \cite{blendin2018}.

\vspace{-0.2cm}
\subsection{Predicting overflows on live data}
In the 10-fold experiment we conducted, each trained model predicts 2.5 months ahead successfully. Thus, retraining should occur at most every month. As explained in section \ref{comp_res}, this is feasible.

\vspace{-0.2cm}
\subsection{Automated reaction}
\begin{figure}[ht!]
\centering
\includegraphics[width=6cm]{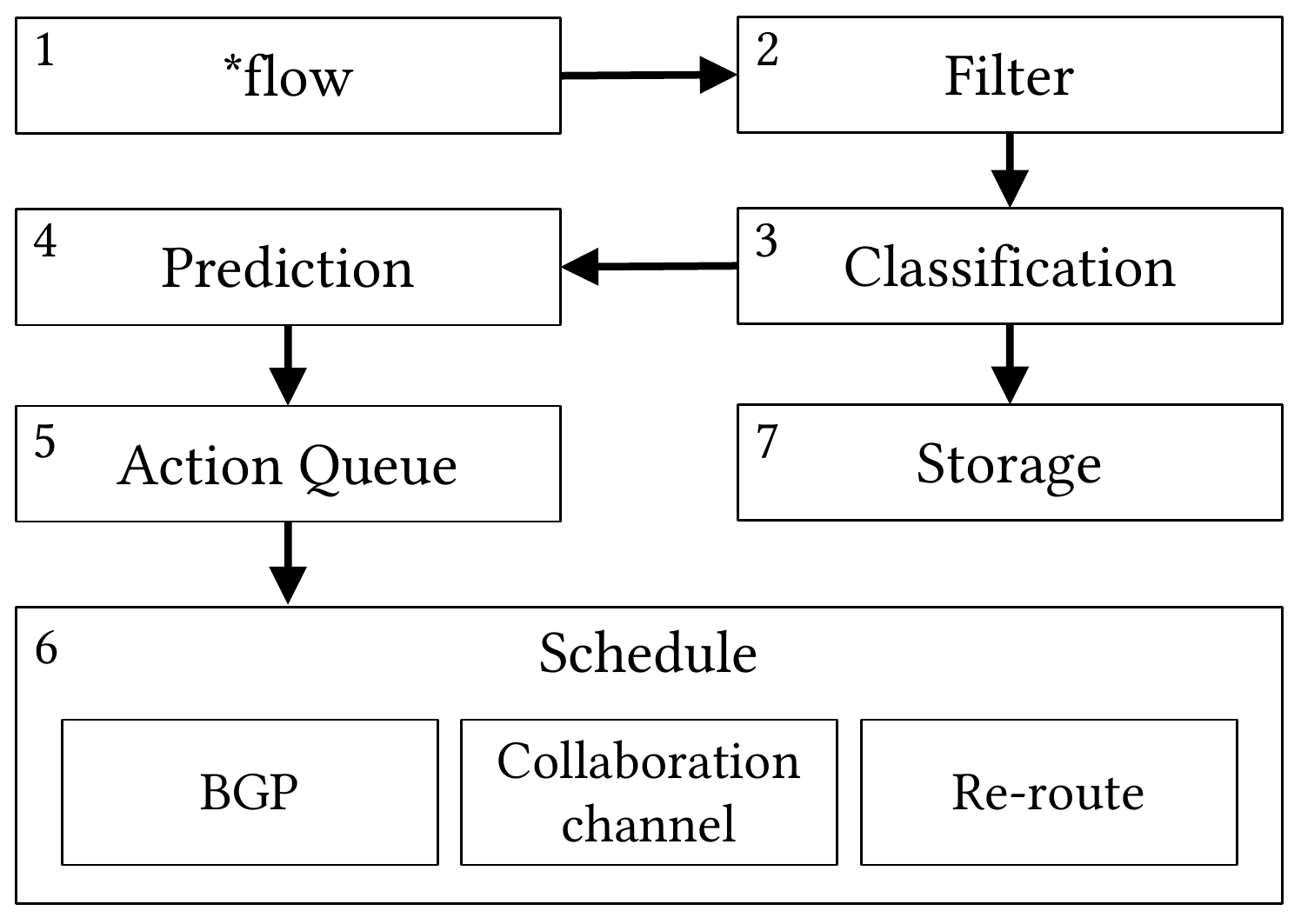}
\caption{Schematic overview of processing live data for automated reactions}
\label{fig:automatedFlow-2}
\vspace{-0.2cm}
\end{figure}

The automated reaction is fed by a continuous stream of data. In the case of overflow prediction, the most valuable source of data is NetFlow and, to some extent BGP. NetFlow is used to identify the amount and source of the traffic that is being ingressed into the Network, while BGP is used to cluster the source IP addresses into autonomous system numbers (ASNs) to reduce the complexity of the data. Figure~\ref{fig:automatedFlow-2} shows a sample processing layout on how this pipeline works which is built upon the architecture of the Collaboration Engine\cite{pujol2019}. The steps in processing are as follows:

\begin{enumerate}
    \itemrange{2} These steps are pre-processing to handle multiple input streams in possible different flow formats (\textit{*flow}), target the correct links (\textit{filter}) and correct for possible double counting as well as correlate with additional data sources like SNMP (\textit{classification}) 
    \item The overflow \textit{prediction} is placed here. Based on the input stream from the classification, the overflows are predicted here. Multiple prediction modules can be run in parallel, making it easy to swap them when re-training.
    \item As each network has different architecture, vendors, and strategy, a generic step is needed to compile the list of actions required to mitigate the overflow. Furthermore, the \textit{action queue} is influenced by the communication channels that are available for general overflow mitigation.
    \item With the given actions compiled, the \textit{schedule} uses different channels to apply the changes to the network. This step is generic and not complete in the schematic, as there are possibly more interaction models available. 
    \item All information fed into the \textit{prediction} model is also stored in long term \textit{storage}. This allows for offline re-training on arbitrary data at a later point in time.
\end{enumerate}

All of the above steps are straight forward, with the exception of the Action Queue and the Schedule of actions towards the Network. 

When dealing with overflows, we first need to take a look at what means of communication between the hyper-giant and the ISP are available. Depending on the scenario, the reactions will be very different. To this end, we define three points that increase the collaboration.
\vspace{-0.2cm}
\subsection{Action Queue}

An ISP has two options to build the action queue depending on mechanisms available for mitigation:

\textit{ISP Internal}.
    An ISP itself has multiple options to influence ingress traffic depending on the hyper-giant's architecture. Possible reactions are (1) re-assigning end-users to different DNS clusters to move users from one mapping region to another when the hyper-giant is using DNS-based mapping. (2) Engineer the internal egress traffic. (3) Modify the external BGP announcement via MED, de-aggregation or revoke to force traffic to shift to another ingress.
\vspace{-0.2cm}
\paragraph{Collaboration Channel}
      The hyper-giant's server selection and the networks need to have a direct communication channel that allows for data exchange about mapping and available capacities before an overflow is happening.
      
      There are multiple protocols capable of transporting this information. Most commonly, (4) mapping information is transported via BGP, either in MED, by prefix de-aggregation. However, doing this via a BGP channel automatically combines the reachability with the connectivity. While these two can overlap, the reachability of a prefix is orthogonal to the mapping infrastructure. To separate the two, another option is (5) out-of-band BGP sessions. However, BGP was not designed to transport multi-dimensional information or correlate prefixes (i.e. obtain the value of a certain prefix in relation to another prefix) other than possibly pushing this into the communities. But communities are ``only'' a 32-bit field. This space can be overloaded with the complexity of the data.
      
      When an out-of-band channel is used, it is advantageous to use a protocol designed for this, i.e. ALTO \cite{rfc7285}. It runs out of band, is independent of BGP, and can correlate arbitrary groups of location identifiers with preferences.

Note that, for the same ISP, it might need to run different solutions for traffic steering depending on the architecture and collaboration channel available to different hyper-giants. Once the action queue is built, this should be applied to the network and/or the Collaboration channel to take effect.

\vspace{-0.2cm}
\section{Conclusions}
We presented a traffic overflow prediction technique relying on an ensemble of automatically generated deep neural networks. 
We provided evidence for the robustness and accuracy of our models by applying it to NetFlow data of a large European ISP quantifying traffic emanating from ten hyper-giant ASes. 
Additionally, our model's ability to pin-point the overflowing traffic link is helpful for the goal of avoiding the predicted overflow events. We used regression models to solve both the tasks of predicting overflow traffic volumes, and predicting overflow events.

To further assess long-term validity of the trained overflow prediction models, they were tested on data from the COVID-19 outbreak period. 
As expected the performance of models trained on 2018-2019 data deteriorated during the peak of these extraordinary circumstances.
Nevertheless, a rise in performance was observed when movement limitations around the globe were lifted.

Using deep learning explainability methods, we have shown that the models we trained learned sensible patterns. 
For example, The prediction is affected the most by data prior to the prediction period and from the previous week. 

Additionally, we proposed methods to mitigate overflow events. For future work, we plan to implement these methods in a simulation environment to obtain evidence that real-time overflow handling is achievable.
\label{sec:conclusions}

\bibliographystyle{ACM-Reference-Format}
\bibliography{overflow_prediction}
\end{document}